\documentclass{jfm}

\usepackage{natbib}
\usepackage{amsbsy}
\usepackage{psfrag}
\usepackage{amsmath}
\usepackage{bm}
\usepackage{soul}
\usepackage{verbatim}
\usepackage[dvipsnames]{xcolor}
\usepackage{graphicx}\usepackage{lineno}
\usepackage[export]{adjustbox}
\usepackage{tikz}
\usepackage{booktabs}
\usepackage{multirow}
\usepackage{enumitem}
\usepackage{setspace}

\usetikzlibrary{shapes}
\definecolor{blueblue}{RGB}{0,114,190}
\definecolor{greengreen}{rgb}{0.16, 0.67, 0.53}
\definecolor{brickred}{rgb}{0.7960, 0.2550, 0.3290}

\usepackage{float}
\usepackage{placeins}

\usepackage{url}
\usepackage[colorlinks=true,
             linkcolor=blue,
             urlcolor=blue,
             citecolor=blue]{hyperref}

\usepackage{graphicx,color}
\usepackage{amssymb,latexsym}
\usepackage{url}

\def\drawline#1#2{\raise 2.5pt\vbox{\hrule width #1pt height #2pt}}

\def\trian{\raise 1.25pt\hbox{$\scriptstyle\triangle$}\nobreak}

\def\dtrian{\raise 1.25pt\hbox%
{$\scriptscriptstyle\bigtriangledown$}\nobreak}

\def\squar{\raise 1.25pt\hbox{$\scriptstyle\Box$}\nobreak}

\def\diamon{\raise 1.25pt\hbox{$\scriptstyle\diamond$}\nobreak}

\def\beq{\begin{equation}}
\def\eeq{\end{equation}}

\def\ggg{{\it g}}

\shorttitle{An analytical model for rotors in confined flow across operating regimes} 
\shortauthor{I.M.L. Upfal et al.} 
\title{An analytical model for rotors in confined flow across operating regimes}
\author{
I.M.L. Upfal\aff{1}, K.J. McClure\aff{1}, K.S. Heck\aff{1}, S. Pieris\aff{2}, J.W. Kurelek\aff{2}, M. Hultmark\aff{3},
\and
M.F. Howland\aff{1}\corresp{\email{mhowland@mit.edu}}
}
\affiliation
{
\aff{1}
Civil and Environmental Engineering, Massachusetts Institute of Technology, \\
Cambridge, MA 02139, USA
\aff{2}
Queen's University, Kingston, Ontario, Canada, K7L 3N6 

\aff{3}
Princeton University, Princeton, NJ 08544, USA
}

\begin{document}

\maketitle

\begin{abstract}
Rotors operating in confined flows, or blockage, are commonly encountered in wind and water tunnels, as well as in shallow or dense deployments of hydrokinetic turbines. 
Confinement induces a streamwise pressure gradient in the channel, modifying the rotor induction, thrust, and power.
To account for these effects, physics-based or empirical blockage corrections are used as a transfer function between the dynamics of an object operating in confined and unconfined settings.
However, existing blockage models are largely only applicable to rotors operating at relatively low thrust coefficients, such that the assumptions of classical momentum theory are valid.
Further, rotors are often partially misaligned with the inflow, which modifies both the geometric blockage and the thrust force, whereas existing blockage models assume perfectly aligned flow conditions.
We develop a generalised engineering model for an actuator disk operating in confined flow at arbitrary misalignment angles and thrust coefficients, termed the Unified Blockage Model. 
The analytical model shows excellent agreement with large eddy simulations of an actuator disk and elucidates the coupled interactions between thrust, misalignment, and blockage. 
To predict bladed rotor dynamics, the Unified Blockage Model is incorporated into a blade element momentum (BEM) model framework and validated against blade-resolved simulations across a wide range of tip-speed and blockage ratios. 
Finally, a blockage correction method is developed based on the Unified Blockage Model and validated against a suite of numerical and experimental data.
\end{abstract}

\section{Introduction}
Fast-running predictive models for the performance of turbomachines, including hydrokinetic/wind turbines and propellers, largely rely on momentum theory \cite[]{rankine1865, froude1878, froude1889}.
Turbomachine rotors generate thrust forces that are either opposing (turbines) or in the direction of the flow (propellers). 
Given this thrust force, the velocity at the rotor cross-section differs from the freestream velocity because of the velocity induced by the thrust.
To predict the induced velocity, momentum theory represents the rotor as a porous actuator disk that imparts a uniform thrust force on the surrounding flow.
Originally derived for propellers, momentum theory is now widely used in the design and analysis of hydrokinetic and wind turbines \cite[]{sorensen2011aerodynamic, adcock2021fluid}.
Simple optimisation of the operating conditions (thrust coefficient) using the momentum theory model results in the well-known  Lanchester-Betz-Joukowsky limit \cite[]{van2007lanchester}, which predicts the theoretical maximum efficiency (coefficient of power) for a turbine operating in unconfined flow.
To yield a fully-predictive model of rotor thrust, power, and wake dynamics, the induction predictions from momentum theory can be coupled with a blade element model \cite[]{glauert1935airplane} that predicts the thrust coefficient based on lift and drag forces on rotating hydro/aerofoils.
The lift and drag forces depend on the velocity at the rotor (induction), yielding the blade element momentum (BEM) approach \cite[e.g.,][]{vogel2018blade, madsen2020implementation}. 
Therefore, accurate BEM predictions rely on both a blade element model that predicts blade-level forces given a known velocity field at the rotor, as well as a momentum closure that predicts the velocity field at the rotor considering the geometry and dynamics of the flow. 
However, in many engineering applications, turbomachines operate in flow configurations that are not appropriately described by classical momentum theory due to its simplifying assumptions, geometric arguments, and boundary conditions.

Hydrokinetic turbines used in tidal and river energy harvesting often operate in confined settings where blockage, the ratio between the swept area of the turbine and the channel cross section, is not negligible \cite[]{garrett2005power}.  
Similarly, in experimental wind tunnels \cite[]{barlow1999low} and water flumes \cite[]{ross2020experimental}, blockage can significantly affect the dynamics and performance of turbomachines \cite[]{chen2011blockage}.
A parallel set of research seeks to investigate the influence of blockage at the scale of arrays of hydrokinetic turbines, which can be parsed separately from the rotor (local) blockage effects \cite[]{nishino2012efficiency, nishino2013two}.
More recently, modelling efforts for arrays of wind turbines have investigated the wind farm blockage effect \cite[]{bleeg2018wind}, a phenomenon distinct from the local/global blockage for hydrokinetic turbines where the mean flow speed in the atmospheric boundary layer reduces upwind and within large wind turbine arrays.
Although the swept area of wind turbines is often much smaller than the cross-section of the atmospheric boundary layer, shallow boundary layer heights and large, dense wind farms can trigger significant blockage effects and associated pressure gradients \cite[]{ndindayino2025effect}.

Meanwhile, turbomachines often operate in non-ideal settings where the mean inflow is not aligned with the rotor.
Such circumstances are common for hydrokinetic/wind turbines because of the chaotic, turbulent environmental flow in which they operate, along with slowly reacting yaw controllers \cite[]{fleming2014field, frost2015effect, frost2017impact, piano2017tidal}.
Recent attention has also focused on wake steering, where the wake is laterally deflected by intentional yaw misalignment, to increase the performance of turbine arrays \cite[]{fleming2019initial, howland2022collective, meyers2022wind}.
Similarly, propellers frequently operate in misalignment, often called drift \cite[]{di2014wake}.

Classical momentum theory does not account for the confined and misaligned settings described above, which can yield significant error in predictive models.
The first-order effects of blockage and inflow-rotor misalignment are to generate pressure gradients and lateral velocities, respectively, which are not modelled in classical momentum theory.
Substantial previous effort has been made to model the effect of blockage on rotor dynamics, starting with \cite{glauert1935airplane}, which is based on a linearisation in the low blockage regime.
More recent studies \cite[]{mikkelsen_sorensen2002, garrett2007efficiency, bahaj2007power, werle2010wind, segalini2014confinement, vogel2018blade, ndindayino2025effect} have developed improved models for the effects of flow confinement on rotor dynamics.
A further limitation of existing blockage models and blockage corrections that are extensions of classical momentum theory is that they cannot predict the dynamics of high thrust rotors \cite[]{ross2020experimental}, because momentum theory is not applicable for high thrust rotors due to the assumption that the wake pressure recovers back to freestream pressure \cite[]{roshko1955wake}. 
This limitation has been recently addressed in unconfined flows~\cite[]{steiros2018drag, liew2024unified}.
Building on the model proposed in \cite{steiros2018drag}, \cite{steiros2022analytical} proposed an extension to simultaneously account for high thrust rotors and confined flow, but it assumes perfect alignment between the rotor and the inflow. 
Separately, recent efforts have focused on modelling rotors operating in misalignment with the inflow \cite[]{bastankhah2016experimental, shapiro2018modelling, heck2023modelling, tamaro2024power}.
In parallel, \cite{liew2024unified} developed an extension of classical momentum theory for unconfined flow that yields low error across the thrust coefficients and rotor-inflow misalignment angles relevant to wind turbines, but it does not capture confinement effects.
To summarise, there is a need for a generalised momentum model which can address blockage, rotor-inflow misalignment, and arbitrary thrust coefficients of realistic rotors in a unified framework for applications including blade element momentum engineering modelling and blockage corrections.
Such a model should both accurately predict key quantities at the rotor, such as the induction, thrust, and power, and should also accurately predict the wake and bypass velocities and pressure that are consistent with conservation of mass, momentum, and energy.

In this study, we develop a generalised momentum theory that accounts for blockage, rotor-inflow misalignment, and arbitrary thrust coefficients.
The analytical model, hereafter referred to as the Unified Blockage Model, is validated against large eddy simulations (LES) of an actuator disk across 80 independent combinations of blockage ratios, thrust coefficients, and misalignment angles.
We further incorporate the Unified Blockage Model within a blade element momentum (BEM) framework to predict the performance of bladed rotors under arbitrary blockage and misalignment, and validate the results against blade-resolved simulations.
Finally, we present a new blockage correction method, which is a widely used approach for mapping rotor measurements between blockage ratios, based on the Unified Blockage Model.
The blockage correction method is evaluated in comparison to computational fluid dynamics simulations as well as experiments of an axial flow hydrokinetic turbine.

The remainder of this paper is organised as follows. 
The Unified Blockage Model, a generalised momentum model for confined rotors, is derived and integrated into BEM and blockage correction methods in \S\ref{sec:model_overview}.
The LES methodology used to validate the Unified Blockage Model is described in \S\ref{sec:les_setup}. 
We present the results of the model predictions in \S\ref{sec:results}.
Finally, the key findings of the work are summarised in \S\ref{sec:conclusions}.

\section{A generalised framework for modelling rotors in blockage}
\label{sec:model_overview}
In this section, we derive the Unified Blockage Model in \S\ref{sec:model_derivation}, which is a generalisation of classical momentum theory that accounts for arbitrary thrust coefficients, rotor-inflow misalignment, and blockage. 
Next, in \S\ref{sec:bem_model}, the Unified Blockage Model is coupled with a blade element closure yielding a fully predictive model for bladed rotors, termed the Unified BEM model. 
Finally, a method for applying the Unified Blockage Model to account for blockage in measurements of rotor thrust and power (a blockage correction) is developed in \S\ref{sec:correction_method}.

\subsection{A blockage model for misaligned actuator disks across thrust coefficients}
\label{sec:model_derivation}

\begin{figure}
    \centering
    \includegraphics[width=0.8\linewidth]{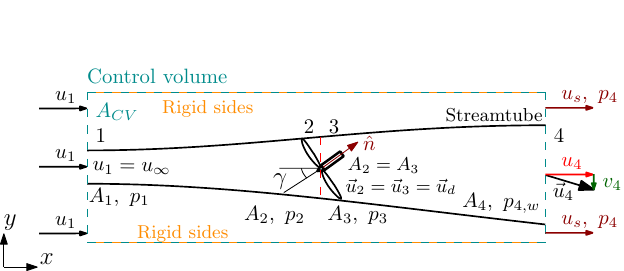}
    \caption{Control volume for analysis of a confined, misaligned turbine.
    The streamwise and spanwise directions are $x$ and $y$, respectively.
    The turbine is yaw/tilt misaligned at angle $\gamma$, where positive misalignment is a counter-clockwise rotation.
    The sides of the control volume are rigid so there is no mass flux through them.
    As in classical momentum theory, we consider four stations for the analysis, and the flow variables are labeled with the corresponding station as subscript numbers.
    The cross-sectional areas, streamwise velocities, spanwise velocities, and pressures are denoted as $A$, $u$, $v$, and $p$, respectively.
    The pressure at the outlet is $p_{4,w}$ within the streamtube and $p_{4}$ outside of the streamtube.
    The unit vector normal to the misaligned turbine is $\hat{n}$.
    }
    \label{fig:cv}
\end{figure}

We aim to develop an analytical model for the induction, thrust, and power of an actuator disk accounting for flow confinement (blockage), misalignment between the inflow and rotor, and arbitrary thrust coefficients, including high-thrust operation.
The non-dimensional parameter describing the rotor-normal, rotor-averaged induced velocity is \cite[]{heck2023modelling}
\begin{equation}
    a_n = 1 - \frac{ \vec{u}_d \cdot \hat{n} }{ u_{\infty} \cos(\gamma)},
    \label{eq:an}
\end{equation}
which is the generalisation of the standard (axial) induction factor in classical one-dimensional momentum modelling \cite[]{van2007lanchester}.
The rotor-averaged velocity at the rotor disk is $\vec{u}_d$, the freestream velocity is $u_\infty$, and the misalignment angle between the inflow and the rotor-normal vector $\hat{n}$ is $\gamma$.
To derive a model for the induction factor $a_n$, it is convenient to describe the rotor aerodynamics (i.e., blade aerofoil properties, number of blades, etc.) and control (tip-speed ratio, blade pitch angle, etc.) with a single parameter, the local thrust coefficient $C_T^\prime$.
The thrust force for the actuator disk is thus defined as \cite[]{calaf2010large}
\begin{equation}
    \vec{F}_T = -\frac{1}{2} \rho C_T^{\prime} A_d (\vec{u}_{d} \cdot \hat{n})^2 \hat{n},
\label{eq:ft_full_les}
\end{equation}
where $\rho$ is the fluid density and $A_d = \pi D^2/4$ is the rotor area, with $D$ the rotor diameter.
We will explicitly connect the value of $C_T^\prime$ to the rotor design and control properties in \S\ref{sec:bem_model}. 

We now consider analysis of the control volume shown in Figure~\ref{fig:cv}, which has rigid sides that are slip-walls. 
This is appropriate for wind \cite[]{mikkelsen_sorensen2002} and water \cite[]{ross2020experimental} tunnels and also for hydrokinetic turbine settings in low Froude number regimes \cite[]{garrett2007efficiency}.
The flow is assumed to be inviscid (frictionless), and the inflow is uniform.
The freestream velocity and pressure are denoted $u_\infty$ and $p_1$, respectively. 
The velocity and pressure at the outlet outside of the streamtube are $u_s$ and $p_4$, respectively. 
To allow for the additional pressure deficit within the wake associated with base suction effects at high thrust~\cite[]{liew2024unified}, the outlet pressure inside the streamtube is defined as $p_{4,w}$.
The streamwise and lateral wake velocities at the streamtube outlet are $u_4$ and $v_4$, respectively.
Conservation of streamwise momentum in the control volume yields
\begin{equation}
\label{eq:mom_cons}
\vec{F}_T \cdot \hat{i} = \rho u_4^2 A_4 + \rho u_s^2(A_{CV}-A_4) - \rho u_\infty^2 A_{CV} + A_{CV} (p_4 - p_1) + A_4(p_{4,w} - p_4),
\end{equation}
where $A_4$ is the area of the streamtube at station 4, and $A_{CV}$ is the cross-sectional area of the control volume in the $yz$-plane.
Conservation of energy inside the streamtube implied by the Bernoulli equation yields~\cite[]{liew2024unified}
\begin{equation}
a_n = 1 - \sqrt{\frac{u_\infty^2 - u_4^2 - v_4^2}{C_T^\prime \cos^2(\gamma) u_\infty^2}-\frac{(p_{4,w} - p_1)}{\frac{1}{2} \rho C_T^\prime \cos^2(\gamma) u_\infty^2}}.
\end{equation}
The Bernoulli equation applied along a streamline outside the streamtube that is inside that control volume is
\begin{equation}
\frac{1}{2} \rho u_\infty^2 + p_1= \frac{1}{2} \rho u_s^2 + p_4,
\label{eq:mass_1}
\end{equation}
where $u_s$ is the bypass velocity that can be larger than $u_1$ due to the flow confinement.
Mass conservation in the streamtube between the actuator disk and outlet gives
\begin{equation}
    u_\infty(1-a_n)\cos(\gamma)A_d =u_4 A_4,
\label{eq:mass_2}
\end{equation}
and mass conservation in the control volume gives
\begin{equation}
 A_{CV} u_\infty = u_4 A_4 + (A_{CV} - A_4) u_s.
\label{eq:bernoulli_2}
\end{equation}
The mass flux out of the control volume at the rigid sides is explicitly zero in this equation. 
Finally, the lateral velocity generated by misalignment is modelled using a lifting line approach \cite[]{shapiro2018modelling, heck2023modelling}
\begin{equation}
\label{eq:lifting-line}
v_4 = -\frac{1}{4} C_T^\prime (1-a_n)^2 \sin(\gamma) \cos^2(\gamma) u_\infty.
\end{equation}


Combining \eqref{eq:mom_cons}-\eqref{eq:lifting-line} yields a system of six coupled nonlinear equations which is the final form of the Unified Blockage Model
{\setlength{\jot}{1em}
\begin{align}
\label{eq:final_1}
a_n &= 1 - \sqrt{\frac{u_\infty^2 - u_4^2 - v_4^2}{C_T^\prime \cos^2(\gamma) u_\infty^2}+\frac{(p_1- p_{4,w})}{\frac{1}{2} \rho C_T^\prime \cos^2(\gamma) u_\infty^2}} \\
\label{eq:final_2}
\frac{u_4}{u_\infty} &= \frac{A_d}{A_4}(1 - a_n) \cos(\gamma) \\
\label{eq:final_3}
\frac{v_4}{u_\infty} &= -\frac{1}{4} C_T^\prime (1-a_n)^2 \sin(\gamma) \cos^2(\gamma) \\
\label{eq:final_4}
\frac{u_s}{u_\infty} &= 1 + \frac{\beta A_4 (u_\infty - u_4)}{(A_d - \beta A_4)u_\infty}\\
\label{eq:final_5}
\frac{A_4}{A_d} &= \frac{\frac{1}{2}C_T'(1-a_n)^2 \cos^3(\gamma) u_\infty^2 + (1/\beta)(u_s^2 - u_\infty^2 - (p_1 - p_4)/\rho)}{(p_4 - p_{4,w})/\rho + u_s^2 - u_4^2} \\
\label{eq:final_6}
\frac{p_1 - p_4}{\rho u_\infty^2} &= \frac{1}{2} (u_s^2 / u_\infty^2 - 1)
\end{align}}
where $\beta = A_d/A_{CV}$ is the geometric blockage ratio.
The Unified Blockage Model comprises six equations, but contains seven unknowns. 
Specifically, the wake pressure $p_{4, w}$ is the unclosed term. 
\cite{liew2024unified} developed a model for the base suction pressure deficit in the wake of a turbine operating in unconfined flow. 
Based on our suite of large eddy simulations in this study, we observe that the wake pressure becomes homogeneous across the entire outlet ($p_{4,w}=p_4$) in the limit of high blockage. 
Specifically, the pressure difference between the wake and bypass flows becomes negligible compared to the much larger pressure gradient resulting from blockage. 
However, the wake pressure $p_{4,w}$ significantly departs from the outer pressure $p_4$ for low blockage and high thrust.
As such, we model the pressure difference between wake and bypass flows as $p_{4,w} - p_4 = (1 - \beta)p_\textrm{UMM}(C_T', \gamma)$, where $p_\textrm{UMM}$ is the pressure deficit given by the Unified Momentum Model~\cite[]{liew2024unified} for unconfined flows. 
This pressure model captures the departure of the wake pressure from the outer pressure in the low blockage regimes ($\beta\rightarrow0$) and captures the convergence of the wake and outer pressure for higher blockage values.
For completeness, a description of the Unified Momentum Model's pressure closure is included in Appendix~\ref{sec:appendix_umm}. 
This formulation yields high accuracy across $\beta$ and $C_T'$ by yielding the correct limiting behaviour for high thrust and low blockage, and returning to a constant wake pressure for high blockage. 
See \S\ref{sec:results_les} for validation of the pressure closure against LES.

As described previously, the local thrust coefficient $C_T^\prime$ is a useful input quantity since it compactly represents the rotor design and control settings.
Further, $C_T^\prime$ is independent of the misalignment angle $\gamma$ and the blockage ratio $\beta$.
Often, rotors are characterized by a thrust coefficient $C_T$, which is defined as $C_T = 2 {\|\vec{F}_T\|} / (\rho A_d u_\infty^2),$ where ${\|\vec{F}_T\|}$ is the magnitude of the thrust force. 
The above equations \ref{eq:final_1}--\ref{eq:final_6} describing the Unified Blockage Model can be closed when $C_T$ is used as the input thrust coefficient variable instead of $C_T^\prime$ by adding an additional equation to the set above
\begin{equation} 
    \label{eq:final_7}
    C_T' = \frac{C_T}{(1-a_n)^2\cos^2(\gamma)}.
\end{equation} 
Note that the thrust coefficient $C_T$ depends on the local thrust coefficient $C_T^\prime$, the misalignment angle $\gamma$, and the blockage ratio $\beta$ (the effect of the blockage ratio comes through the induction factor $a_n$). 
To summarize, the model solved using $C_T'$ as the input thrust variable and Eqs.~\ref{eq:final_1}--\ref{eq:final_6} can be written compactly as
\begin{equation}
\label{eq:local_thrust_model}
[a_n,\; u_4/u_\infty,\; v_4/u_\infty,\; u_s/u_\infty,\; A_4/A_d,\; (p_1-p_4)/\rho u_\infty^2]
= f_\textrm{Unified}(C_T', \gamma, \beta)
\end{equation}
and will be referred to as the $C_T'$-formulation.
When using $C_T$ as the input thrust variable together with Eqs.~\ref{eq:final_1}--\ref{eq:final_7}, the model can be written as
\begin{equation}
\label{eq:thrust_model}
[a_n,\; u_4/u_\infty,\; v_4/u_\infty,\; u_s/u_\infty,\; A_4/A_d,\; (p_1-p_4)/\rho u_\infty^2,\; C_T']
= f_\textrm{Unified}(C_T, \gamma, \beta)
\end{equation}
and will be referred to as the $C_T$-formulation. Here, $f_\textrm{Unified}$ denotes the converged solution to the appropriate set of equations.

The nonlinear coupled system is solved using a standard root-finding algorithm. 
As with most such methods, fast and reliable convergence depends on providing a suitable initial guess.
When solved with the $C_T'$-formulation, initial guesses for induction $a_n$ and wake velocities $u_4/u_\infty$ and $v_4/u_\infty$ are obtained directly from the Unified Momentum Model~\cite[]{liew2024unified} which is the limit of $\beta\rightarrow0$. 
These are combined with $
A_4/A_d = (1 - a_n)\cos(\gamma)(u_\infty/u_4)$,
together with Eqs.~\ref{eq:final_4} and~\ref{eq:final_6} to determine $u_s/u_\infty$ and $(p_1 - p_4)/\rho u_\infty^2$. 
Since the initial guess is given by the $\beta\rightarrow0$ limit, this initial guess can be too far from the final solution for large values of $\beta$.
So for larger values of $\beta$ (e.g. $\beta=0.3$), we first solve the model for $\beta=0.1$ based on initialization from the $\beta=0$ limiting case as described above, then we solve the model at $\beta=0.2$ with initialization from the $\beta=0.1$ results and so on until the solution at the desired $\beta$ is reached.
When solved using the $C_T$-formulation, initial guesses are generated via pre-tabulation. 
First, the $C_T'$-formulation is solved across a range of $C_T'$, $\gamma$, and $\beta$ which gives the corresponding values of $C_T$ using Eq.~\ref{eq:final_7}. 
Then, the desired initial guesses for the model outputs
$
(a_n,\; u_4/u_\infty,\; v_4/u_\infty,\; u_s/u_\infty,\; A_4/A_d,\; (p_1 - p_4)/\rho u_\infty^2,\; C_T')$
are tabulated as a function of the input variables $(C_T, \gamma, \beta)$. 
Initial guesses for any combination of input variables can then be obtained using linear interpolation.

In the limit of zero misalignment and low thrust coefficients, the Unified Blockage Model returns to the actuator disk model developed by \cite{garrett2007efficiency} for confined flow, while in the limit of zero blockage ($\beta\rightarrow 0$), the model yields the Unified Momentum Model for unconfined flows developed by~\cite{liew2024unified}. As will be shown in~\S\ref{sec:results_les}, nonlinear interactions arise between blockage, thrust coefficient, and misalignment angle, such that the previous models cannot be superposed and instead must be modelled jointly, as the present approach provides. 

\subsection{A generalised blade element momentum model for rotors in confinement}
\label{sec:bem_model}

The Unified Blockage Model presented in the previous section predicts the induced velocities, and therefore the velocity at the rotor, based on an input blockage ratio, misalignment angle, and thrust coefficient. 
However, the thrust coefficient for an arbitrary rotor design and control strategy is not known a priori. 
Here, a blade element model is used to calculate the thrust coefficient based on the velocity at the rotor, the aerofoil properties, misalignment angle, blade pitch angle, and tip-speed ratio.
The blade element model is solved on a polar grid depending on radial and azimuthal position $(\mu, \psi)$.
These positions are formulated non-dimensionally, with non-dimensional radius $\mu=r/R$, where $R=D/2$ is the rotor radius and $r$ is the radial coordinate, and $\psi$ in radians. 
The tip-speed ratio is $\lambda = R\Omega/u_\infty$, where $\Omega$ is the angular velocity of the rotor.
The axial and tangential velocity components relative to the blades are
\begin{align}
    \label{eq:BEM-start}
    v_n(\mu,\psi) & = (1-a_n)\cos(\gamma)u_\infty                   \\
    \label{eq:bem_vt}
    v_t(\mu,\psi) & = (1 + a')\lambda \mu u_\infty - (1-a_n)\cos(\psi)\sin(\gamma)u_\infty,
\end{align}
assuming uniform inflow, where $a'$ is the tangential induction factor.
The speed $w$ and angle $\phi$ of the flow relative to the blades are
\begin{align}
    \label{eq:bem_w}
    w^2(\mu,\psi)  & = v_n^2(\mu,\psi) + v_t^2(\mu,\psi)                          \\
    \label{eq:bem_phi}
    \phi(\mu,\psi) & = \tan^{-1}\left(\frac{v_n(\mu,\psi)}{v_t(\mu,\psi)}\right).
\end{align}
The lift and drag coefficients, $C_l$ and $C_d$, are evaluated as a function of the local angle of attack $\alpha(\mu,\psi) = \phi(\mu,\psi) - \theta_t(\mu) - \theta_p$, where $\theta_t$ and $\theta_p$ are the blade twist and pitch angles respectively.
Transforming into the rotor frame, 
the rotor-normal and tangential force coefficients are
\begin{align}
    C_{n}(\mu,\psi)     & = \cos(\phi)C_l(\mu, \alpha(\mu,\psi)) + \sin(\phi)C_d(\mu, \alpha(\mu,\psi))  \\
    C_{tan}(\mu,\psi) & = \sin(\phi)C_l(\mu, \alpha(\mu,\psi)) - \cos(\phi)C_d(\mu, \alpha(\mu,\psi)).
\end{align}
Finally, the thrust coefficient is
\begin{equation}
    C_T(\mu, \psi) = \sigma(\mu)C_{n}(\mu,\psi) \frac{w^2(\mu, \psi)}{u_\infty^2}
\end{equation}
with blade solidity $\sigma(\mu) = Bc/2\pi r$, where $B$ is the number of blades and $c(\mu)$ is the blade chord.
To account for reductions in the blade force coefficients due to the effects of tip vortices, we apply a standard tip loss correction factor
\begin{equation}
    F(\mu) = \frac{2}{\pi}\arccos \left[\exp\left(-\frac{B(1-\mu)}{2\mu\sin{\phi}}\right)\right],
\end{equation}
and use a modified axial force as input to the induction model \cite[]{madsen2020implementation}
\begin{equation}
    \label{eq:axial_force_coeff_corr}
    C_T^{corr}(\mu, \psi) = \frac{C_T(\mu, \psi)}{F(\mu)}.
\end{equation}
The rotor-normal and tangential induction factors are initialized with $a_n(\mu, \psi) = 1/3$ and $a'(\mu, \psi)=0$, respectively. At each iteration, Eqs~\ref{eq:BEM-start}-\ref{eq:axial_force_coeff_corr} are solved to update the rotor normal and tangential forces based on the current values of $a_n(\mu, \psi)$ and $a'(\mu, \psi)$. The blade element model is connected to the Unified Blockage Model derived in~\S\ref{sec:model_derivation} using the $C_T$-formulation $f_\textrm{Unified}(C_T^{corr}(\mu, \psi), \gamma, \beta)$ (Eq.~\ref{eq:thrust_model}) to update $a_n(\mu, \psi)$. 
The tangential induction factor is updated using~\cite[]{liew2024unified}
\begin{equation}
    a'(\mu, \psi) = \frac{\sigma(\mu)C_{tan}(\mu,\psi) w^2(\mu, \psi)}{4 \lambda \mu F(\mu) (1-a_n)\cos(\gamma) u_\infty^2}.
\end{equation}
After the calculation of the new rotor-normal and tangential induction factors, $a_n(\mu, \psi)$ and $a'(\mu, \psi)$, respectively, we return back to Eq.~\ref{eq:BEM-start} to restart the loop.
The coupled system is solved using fixed point iteration with relaxation until converged solutions for $a_n(\mu, \phi)$ and $a'(\mu, \psi)$ are reached.
Finally, the output rotor-averaged thrust is calculated by integrating over the rotor area
\begin{equation}
    \bar{C}_T = \frac{1}{\pi}\int_0^1\int_0^{2\pi} \mu\sigma(\mu)C_{n}(\mu,\psi) \frac{w^2(\mu, \psi)}{u_\infty^2} d\psi d\mu,
\end{equation}
and similarly the rotor-averaged power coefficient is
\begin{equation}
    \bar{C}_P =\frac{1}{\pi} \int_0^1 \int_0^{2\pi} \lambda\mu^2 \sigma(\mu)C_{tan}(\mu,\psi) \frac{w^2(\mu, \psi)}{u_\infty^2} d\psi d\mu.
    \label{eq:BEM-end}
\end{equation}

\subsection{A simple blockage correction based on the Unified Blockage Model}
\label{sec:correction_method}
The Unified BEM method presented in the previous section predicts rotor thrust and power for arbitrary rotor design, tip-speed ratio, pitch angles, rotor-inflow misalignment, and blockage ratio. 
This method solves for the rotor-normal induction using the aerofoil lift and drag coefficients as a thrust closure for the Unified Blockage Model. 
To compute the rotor power, the aerofoil data is then used to determine the torque contributions along the blades.
However, if measurements of rotor thrust and power are available at a single blockage ratio, these can be used to predict $C_T$ and $C_P$ for arbitrary $\beta$ without the need for a blade element model, which can be beneficial because the blade element model requires detailed information on the aerofoils. 
This methodology is widely referred to as a blockage correction \cite[e.g.,][]{ross2020experimental}.

We now use the Unified Blockage Model derived in \S\ref{sec:model_derivation} to develop a simple blockage correction method.
While previous work has sought to correct for blockage by identifying a so-called ``equivalent unblocked freestream velocity'' \cite[]{ross2020experimental}, we instead propose an alternative method that leverages the Unified Blockage Model to directly capture the change in rotor disk velocity induced by blockage. 
Specifically, it is useful to normalise $\lambda$, $C_T$, and $C_P$ by the rotor-normal disk velocity, $\vec{u}_d \cdot \hat{n}$, instead of the freestream velocity $u_\infty$. This yields the local tip-speed ratio $\lambda'$, local thrust coefficient $C_T'$, and local power coefficient $C_P'$. Using $\vec{u}_d \cdot \hat{n} = (1 - a_n) \cos(\gamma) u_\infty$ from Eq.~\ref{eq:an} gives
\begin{equation}
\label{eq:prime_normalization}
\lambda'=\frac{\lambda}{(1-a_n)\cos(\gamma)}, \quad
C_T'=\frac{C_T}{(1-a_n)^2\cos^2(\gamma)}, \quad
C_P'=\frac{C_P}{(1-a_n)^3\cos^3(\gamma)}.
\end{equation}
This local description of rotor thrust and power in terms of the disk velocity, $C_T'(\lambda')$ and $C_P'(\lambda')$, decouples the rotor performance, which is specific to every turbine, from the induction, in contrast to $C_T$ and $C_P$ where rotor performance and blockage effects are intermingled.
As a result, we expect that $C_T'(\lambda')$ and $C_P'(\lambda')$ are independent of $\beta$ if the aerofoil properties, $C_l(\mu, \alpha)$ and $C_d(\mu, \alpha)$, yaw angle, and pitch angles do not change. 
A short analysis using the BEM equations is provided in Appendix~\ref{sec:appendix_norm_bem} to show that $C_T'$ and $C_P'$ depend only on $\lambda'$, $\theta_p$, and $\gamma$ when $C_l(\mu, \alpha)$ and $C_d(\mu, \alpha)$ are fixed.  
Blade-resolved simulations are also used to confirm that $C_T'(\lambda')$ and $C_P'(\lambda')$ are constant across a wide range of $\beta$ in \S\ref{sec:results_correction}. 
The invariance of $C_l$ and $C_d$ to confinement is a critical assumption for all blockage corrections, and implications of non-constant $C_l$ and $C_d$, for example due to Reynolds number effects, are further discussed in \S\ref{sec:results_correction}.

Given the local rotor performance scaling, $C_T'(\lambda')$ and $C_P'(\lambda')$, we now apply it to develop a new blockage correction method.
From measurements of $C_T^{\beta_1}$ and $C_P^{\beta_1}$ at an input blockage ratio $\beta_1$ and input tip speed ratio $\lambda^{\beta_1}$, we determine the induction factor $a_n^{\beta_1}$ using the $C_T$-formulation (Eq.~\ref{eq:thrust_model}) of the Unified Blockage Model. 
Using the induction factor at $\beta_1$, we compute $C_T'(\lambda')$ and $C_P'(\lambda')$ with Eq.~\ref{eq:prime_normalization}.
Next, the $C_T'$-formulation (Eq.~\ref{eq:local_thrust_model}) is used to predict the induction factor at an arbitrary output blockage ratio $\beta_2$. 
Finally, the induction at the output blockage ratio is used to determine $\lambda^{\beta_2}$, $C_T^{\beta_2}$ and $C_P^{\beta_2}$ from $C_T'(\lambda')$ and $C_P'(\lambda')$ with Eq.~\ref{eq:prime_normalization}.
Therefore, we have a concise and physically justified blockage correction method to map measurements of thrust and power between different blockage ratios for a given rotor of interest.
This new blockage correction method based on the Unified Blockage Model is performed as follows:
\begin{list}{(\roman{enumi})}{%
  \usecounter{enumi}
  \setlength{\leftmargin}{3em}
  \setlength{\labelwidth}{2em}
  \setlength{\labelsep}{0.75em}
  \setlength{\itemindent}{1pt}
}
    \item Solve the $C_T$-formulation $f_\textrm{Unified}(C_T, \gamma, \beta)$ (Eq.~\ref{eq:thrust_model}) using the thrust coefficient $C_T^{\beta_1}$ measured at blockage ratio $\beta_1$ to determine the induction factor $a_n^{\beta_1}$ 
    \begin{equation}
        \label{eq:correction_step_1}
        [a_n^{\beta_1}, ...] = f_\textrm{Unified}(C_T^{\beta_1}, \gamma, \beta_1).
    \end{equation}

    \item Use $a_n^{\beta_1}$ to compute the local tip-speed ratio $\lambda'$, local thrust coefficient $C_T'$, and local power coefficient $C_P'$ with Eq.~\ref{eq:prime_normalization} from the data measured at $\beta_1$
    \begin{equation}
        \label{eq:correction_step_2}
        \lambda' = \frac{\lambda^{\beta_1}}{(1 - a_n^{\beta_1})\cos(\gamma)},
\
        C_T' = \frac{C_T^{\beta_1}}{(1 - a_n^{\beta_1})^2\cos^2(\gamma)},
\
        C_P' = \frac{C_P^{\beta_1}}{(1 - a_n^{\beta_1})^3\cos^3(\gamma)}.
    \end{equation}
    
    \item Solve the $C_T'$-formulation $f_\textrm{Unified}(C_T',\gamma, \beta)$ (Eq.~\ref{eq:local_thrust_model}) to determine the induction factor $a_n^{\beta_2}$ at the blockage ratio $\beta_2$ 
    \begin{equation}
        \label{eq:correction_step_3.1}
        [a_n^{\beta_2}, ...] = f_\textrm{Unified}(C_T', \gamma, \beta_2).
    \end{equation}

    \item Use $a_n^{\beta_2}$ to determine the output control and performance parameters $\lambda^{\beta_2}$, $C_T^{\beta_2}$, and $C_P^{\beta_2}$ with Eq.~\ref{eq:prime_normalization}
    \begin{align}
        \label{eq:correction_step_4}
        \lambda^{\beta_2} &= \lambda'(1 - a_n^{\beta_2})\cos(\gamma), \\
        C_T^{\beta_2} &= C_T'(1 - a_n^{\beta_2})^2\cos^2(\gamma), \\
        C_P^{\beta_2} &= C_P'(1 - a_n^{\beta_2})^3\cos^3(\gamma).
    \end{align}
\end{list}
To summarise, this new blockage correction procedure is a mapping from one combination of blockage ratio, tip-speed ratio, thrust coefficient, and power coefficient to another
\begin{equation}
    (\beta_1, \lambda^{\beta_1}, C_T^{\beta_1}, C_P^{\beta_1}) \rightarrow (\beta_2, \lambda^{\beta_2}, C_T^{\beta_2}, C_P^{\beta_2})
\end{equation}
where the mapping is performed by enforcing that $C_T'$ and $C_P'$ are constant for a given $\lambda'$, pitch angles, and rotor-inflow misalignment.

These local parameters $C_T'$ and $C_P'$ can also be interpreted as the unknown rotor performance parameters predicted by the blade element model. Therefore, measurements at a single $\beta$ in conjunction with the Unified Blockage Model can be used in place of the blade element model to make predictions at arbitrary $\beta$. This eliminates the need for aerofoil data, simplifying the modelling effort and reducing uncertainty introduced by the aerofoil polars. 

\section{Large eddy simulation numerical setup}
\label{sec:les_setup}
We perform large eddy simulations with Pad{\'e}Ops \cite[]{ghate2017subfilter, howland2020influence,heck2025coriolis}, an open-source, incompressible, pseudo-spectral solver for the filtered Navier--Stokes equations \cite[]{ghate2018padeops}.
Relevant details are provided in this section, and the complete numerical description and validation are available in published studies \cite[e.g.][]{heck2025coriolis}.
The sigma subgrid scale model is used for subgrid scale stresses \cite[]{nicoud2011using}.
The simulations use inflow conditions consistent with the momentum modelling: uniform inflow with zero freestream turbulence. 
This inflow is set by a fringe region \cite[] {nordstrom1999fringe} in the $x$-direction. 
The turbine is placed $x=5D$ from the inlet and centred laterally and vertically in the domain.
No-flux boundary conditions are used at the rigid sides ($y$-direction), and periodic boundary conditions are used in the $x$- and $z$-directions, with the fringe region providing uniform inflow at the inlet.
The simulation domains used for each blockage ratio are described in Table~\ref{tab:blockage_cases}. 
The grid sizes were selected to ensure that the resolution is approximately constant between blockage ratios.

\begin{table}
\centering
\caption{Description of large eddy simulation domains used for each blockage ratio. The domain lengths and number of grid points in the $x$, $y$, and $z$ directions are denoted $L_x$, $L_y$, and $L_z$, and $N_x$, $N_y$, and $N_z$, respectively.}
\begin{tabular}{ccccc}
\toprule
\textbf{Blockage ratio} & \textbf{$L_x/D$} & \textbf{$L_y/D$} & \textbf{$L_z/D$} & \textbf{$N_x \times N_y \times N_z$ } \\
\midrule
$\beta = 0.005$ & $25$ & $12.5$ & $12.5$ & $640 \times 800 \times 800$ \\
$\beta = 0.1$ & $25$ & $2.80$ & $2.80$ & $640 \times 180 \times 180$ \\
$\beta = 0.2$ & $25$ & $1.98$ & $1.98$ & $640 \times 128 \times 128$ \\
$\beta = 0.3$ & $25$ & $1.50$ & $1.50$ & $640 \times 128 \times 128$ \\
\bottomrule
\end{tabular}
\label{tab:blockage_cases}
\end{table}

The filtered actuator disk model (ADM) methodology developed by~\cite{shapiro2019filtered} is used to model the rotor.
The thrust force depends on the local thrust coefficient $C_T^\prime$ and the velocity at the disk $\vec{u}_d \cdot \hat{n}$ \cite[]{calaf2010large, shapiro2019filtered, heck2023modelling}. 
The actuator disk power is defined as $P = -\vec{F}_T \cdot \vec{u}_d$.
For the inflow-aligned ($\gamma=0$) $C_T'$ sweep, a filter width of $\Delta/D = 3.2\times10^{-2}$ and a disk thickness of $s = 3\Delta x/2$ are used, following~\cite{heck2023modelling}. 
This narrow filter width minimises distortion of the induction, which occurs as a result of smoothing the ADM forcing through filtering. 
The filtered actuator disk correction factor from \cite{shapiro2019filtered} is turned off since it is not needed because of the narrow filter width used.
Additionally, since the correction factor is based on classical momentum theory, which is not valid for the confined case, it would introduce error to the induction of the confined rotor.
To simulate misaligned turbines in unconfined flow, \cite{heck2023modelling} rotated the angle of the inflow while keeping the actuator disk aligned with the computational grid. 
In the present study, simulations of misaligned rotors in a confined channel require the rotor itself to be rotated in order to maintain no-flux boundary conditions at the channel walls. 
This introduces a trade-off between maintaining a sufficiently narrow filter and ensuring a smooth spatial distribution of the turbine forcing for arbitrary misalignment angles. 
To balance these requirements, we employ a grid resolution approximately twice as fine in each spatial direction as \cite{heck2023modelling}, a filter width of $\Delta/D = 6.5\times10^{-2}$, and a disk thickness of $s = 3\Delta x$ for the misaligned cases. Although there is a small sensitivity to the numerical setup, we find this to be at least an order of magnitude smaller than the effect of blockage.
This configuration maintains a relatively narrow filter with minimal influence on induction while ensuring a smoothly distributed forcing for the misaligned ADM.


Simulations are run for four flow-through times $L_x / u_\infty$, with the velocity and pressure fields averaged over the final two flow-throughs, which is sufficient for convergence of first-order statistics in these zero freestream turbulence inflow cases \cite[]{heck2023modelling}.
Outlet quantities within the wake ($u_4$, $v_4$, and $p_{4,w}$) are extracted from the LES using a streamtube seeded along the outer radius of the misaligned ADM ($R_s=0.5D$). 
The near-wake length $x_0$ is defined as the streamwise location at which the minimum $yz$-averaged streamwise velocity occurred within the streamtube (i.e., the location where turbulent wake recovery begins). 
The outlet wake velocity $u_4$ and wake pressure \(p_{4,w}\) are taken as the corresponding $yz$-averaged values at $x_0$ within the streamtube, and the outlet wake area \(A_4\) is defined as the cross-sectional area of the streamtube at this location.
Following~\cite{shapiro2018modelling}, the spanwise initial wake velocity \(v_4\) is defined as the maximum $yz$-averaged spanwise velocity within the wake streamtube over \(0 \le x/D \le 5\). 
To estimate the bypass flow quantities, \(u_s\) and \(p_4\), a separate streamtube of radius \(R_b = 0.55D\) is seeded at the turbine location. 
The streamwise velocity and pressure are then averaged over the channel cross-section outside this streamtube at \(x_0\) to obtain \(u_s\) and \(p_4\), respectively.
Streamtube-averaged LES results for $\beta = 0.005$ are shown for $C_T \leq 1.1$ due to the transition to unsteadiness in the unconfined case~\cite[]{sorensen1998analysis, liew2024unified}.

\FloatBarrier

\section{Results}
\FloatBarrier
\label{sec:results}
This section presents a comparison between the Unified Blockage Model and LES, followed by validation of the Unified BEM and blockage correction methods.
The results of the Unified Blockage Model, derived in \S\ref{sec:model_derivation}, and LES of an actuator disk are presented in \S\ref{sec:results_les}. 
In \S\ref{sec:results_bem}, the Unified BEM model developed in~\S\ref{sec:bem_model} is validated against blade-resolved simulations. 
In \S\ref{sec:results_correction}, the new blockage correction method presented in~\S\ref{sec:correction_method} is validated by mapping rotor thrust and power measurements from blade-resolved simulations, LES of an actuator line model, and experimental measurements across a wide range of tip-speed and blockage ratios.

\subsection{Analytical model of an actuator disk in misalignment and blockage}
\label{sec:results_les}
\begin{figure}
    \centering
    \includegraphics[width=\linewidth]{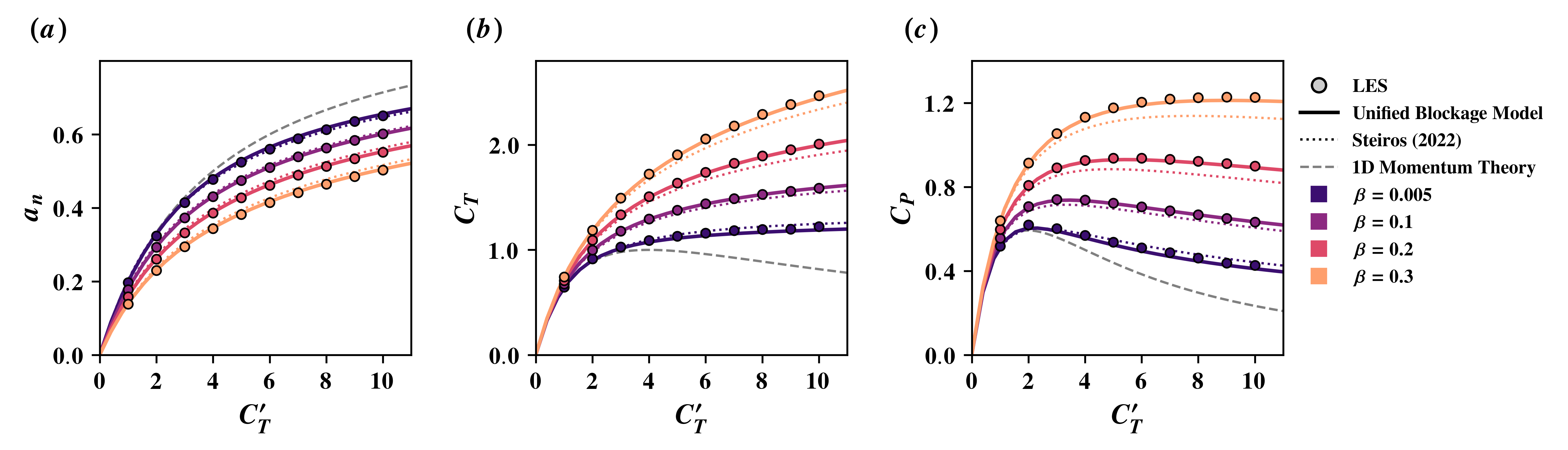}
	\caption{Comparison between the Unified Blockage Model (Eqs.~\ref{eq:final_1}-\ref{eq:final_6}) and LES for (a) rotor-normal induction, (b) thrust coefficient, and (c) power coefficient across local thrust coefficients $C_T'$ and blockage ratios $\beta$ with rotor-aligned inflow ($\gamma=0$). The predictions of classical one-dimensional momentum theory and the blockage model of~\cite{steiros2022analytical} are plotted for reference.}
	\label{fig:ctp_sweep}
\end{figure}

We begin by comparing the analytical model predictions with LES results for an inflow-aligned actuator disk ($\gamma = 0$), before considering misaligned cases.
The induction factor $a_n$, thrust coefficient $C_T$, and power coefficient $C_P$ from LES and the Unified Blockage Model are shown in figure~\ref{fig:ctp_sweep}. 
The analytical model exhibits low error, compared to the LES results, across a wide range of blockage ratios and local thrust coefficients. 
In particular, the present model accurately represents the rotor dynamics in the limit of low blockage and high thrust, where classical momentum theory is not valid.
Moreover, the Unified Blockage Model demonstrates improved agreement with LES compared to the recent model developed by \cite{steiros2022analytical}, especially for higher $\beta$ and $C_T'$.
Figure~\ref{fig:ctp_sweep}($a$) shows that the induction factor $a_n$ decreases with increasing blockage ratio $\beta$.
Consequently, the thrust and power coefficients, which scale as $C_T \sim (1-a_n)^2$ and $C_P \sim (1-a_n)^3$, respectively, both increase with blockage ratio~(Figure~\ref{fig:ctp_sweep}($b,c$)). 

\begin{figure}
    \centering
    \includegraphics[width=0.9\linewidth]{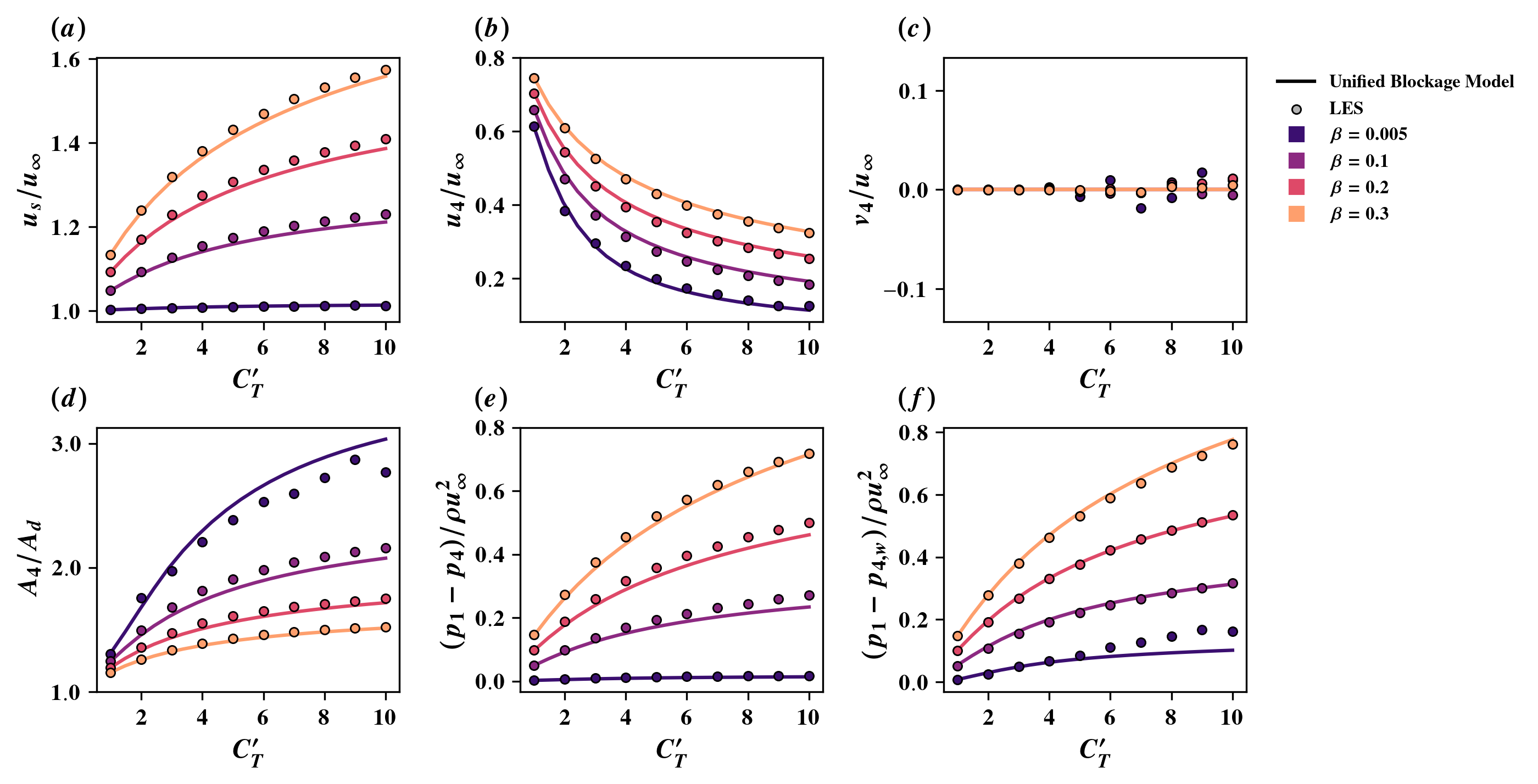}
	\caption{Comparison of (a) bypass velocity, (b) initial streamwise wake velocity, (c) spanwise initial wake velocity, (d) outlet wake area, (e) bypass pressure drop, and (f) streamtube pressure drop between the Unified Blockage Model developed in~\S\ref{sec:model_derivation} and LES of an actuator disk for various local thrust coefficients $C_T'$ and rotor-aligned inflow ($\gamma = 0$).}
	\label{fig:ctp_streamtube}
\end{figure}

In figure~\ref{fig:ctp_streamtube}, the bypass velocity, streamwise and spanwise wake velocities, outlet wake area, and outlet pressures are compared between LES and the Unified Blockage Model. 
Overall, the model predictions show strong agreement with the LES measurements. 
Figure~\ref{fig:ctp_streamtube}($a$) shows that the bypass velocity increases with blockage ratio to satisfy global mass conservation within the channel. Confinement also increases the streamwise wake velocity (figure~\ref{fig:ctp_streamtube}($b$)) as the induction factor is reduced. Consequently, the wake expands less, leading to a smaller outlet wake area at higher blockage ratios (figure~\ref{fig:ctp_streamtube}($d$)).
Figure~\ref{fig:ctp_streamtube}($e$) shows that rotor confinement induces a favourable pressure gradient, $p_1 - p_4$, due the acceleration of the bypass flow around the disk. 
In the absence of blockage, the pressure gradient outside of the streamtube, $p_1 - p_4$, is zero while base suction causes the change in pressure within the streamtube, $p_1 - p_{4,w}$, to be non-zero (figure~\ref{fig:ctp_streamtube}($f$)). The model captures the rotor behaviour in the limit of high thrust and low blockage, where classical momentum theory breaks down, by incorporating the base suction pressure deficit predicted by the Unified Momentum Model for unconfined flows \cite[]{liew2024unified}.
As $\beta$ increases, the pressure gradient increases across the entire channel and dominates the contribution of base suction ($p_1 - p_4 \gg p_4 - p_{4,w}$). Figure~\ref{fig:ctp_streamtube}($f$) demonstrates that the model captures the combined blockage and base suction pressure contributions across a wide range of $\beta$ and $C_T'$.

The blockage-induced favourable pressure gradient, $p_1 - p_4$, allows additional momentum to be extracted from the pressure field.
As a result, the disk velocity increases and the induction factor decreases.
The blockage-induced pressure gradient increases with rotor thrust, such that blockage effects become more pronounced at higher $C_T'$.
This is consistent with the limit of zero thrust coefficient, where there is no favourable pressure gradient from geometric confinement and therefore no blockage effects.
Thus, the blockage effect depends not only on the geometric blockage ratio $\beta$ but also on its coupling with the thrust coefficient ($C_T$ or $C_T^\prime$).

\begin{figure}
    \centering
    \includegraphics[width=\linewidth]{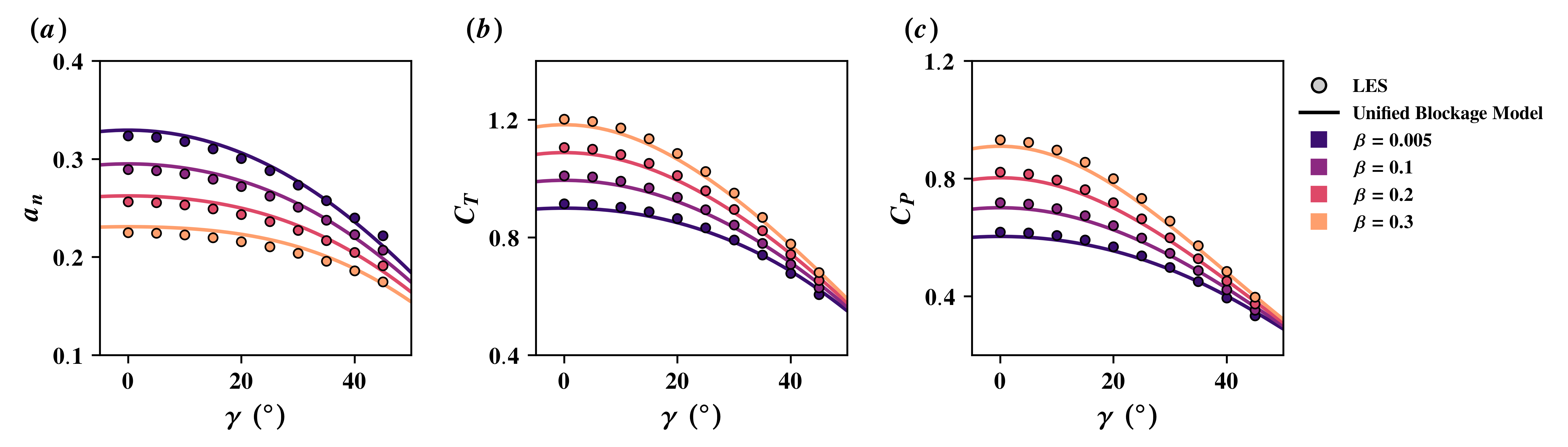}
	\caption{Comparison between the Unified Blockage Model (Eqs.~\ref{eq:final_1}-\ref{eq:final_6}) and LES for (a) induction factor, (b) thrust coefficient and (c) power coefficient across misalignment angles $\gamma$ and blockage ratios $\beta$ with $C_T'=2$. Classical momentum theory and the blockage model of~\cite{steiros2022analytical} are not shown as they assume perfect alignment between the rotor and inflow.}
	\label{fig:tilt_sweep}
\end{figure}

Having established the rotor-aligned performance, the Unified Blockage Model is next assessed under rotor-misaligned inflow. 
In figure~\ref{fig:tilt_sweep}, model predictions of the rotor-normal induction factor $a_n$, thrust coefficient $C_T$, and power coefficient $C_P$ across misalignment angles $\gamma$ and blockage ratios $\beta$ are compared with LES for $C_T' = 2$. 
For an aligned rotor in unconfined flow, $C_T^\prime=2$ results in $C_T=8/9$.
Excellent agreement between the LES and the Unified Blockage Model is observed across all misalignment angles and blockage ratios.
We note that predictions from the model of \cite{steiros2022analytical} are not shown since that study and model only consider perfect alignment between the turbine and inflow. 
In figure~\ref{fig:tilt_sweep}($a$), induction decreases with increasing magnitude of misalignment and increasing blockage. 
As a result of the decreasing induction, the rotor thrust and power coefficients increase with increasing blockage (figure~\ref{fig:tilt_sweep}($b,c$)). 
As the misalignment angle increases, both thrust and power decrease, despite the reduction in induction.
The quantitative outcome of the reduction in the thrust and power with the misalignment angle will depend on the competition between the geometric reduction in the rotor-normal component of the freestream flow velocity against the reduction in induction associated with the misalignment.
The competition between the geometry and the response of the induction factor is why the thrust and power for a misaligned actuator disk are not accurately modelled by geometry alone (i.e. modelling thrust and power reductions using $\cos^2(\gamma)$ and $\cos^3(\gamma)$, respectively, will lead to errors). 

\begin{figure}
    \centering
    \includegraphics[width=0.9\linewidth]{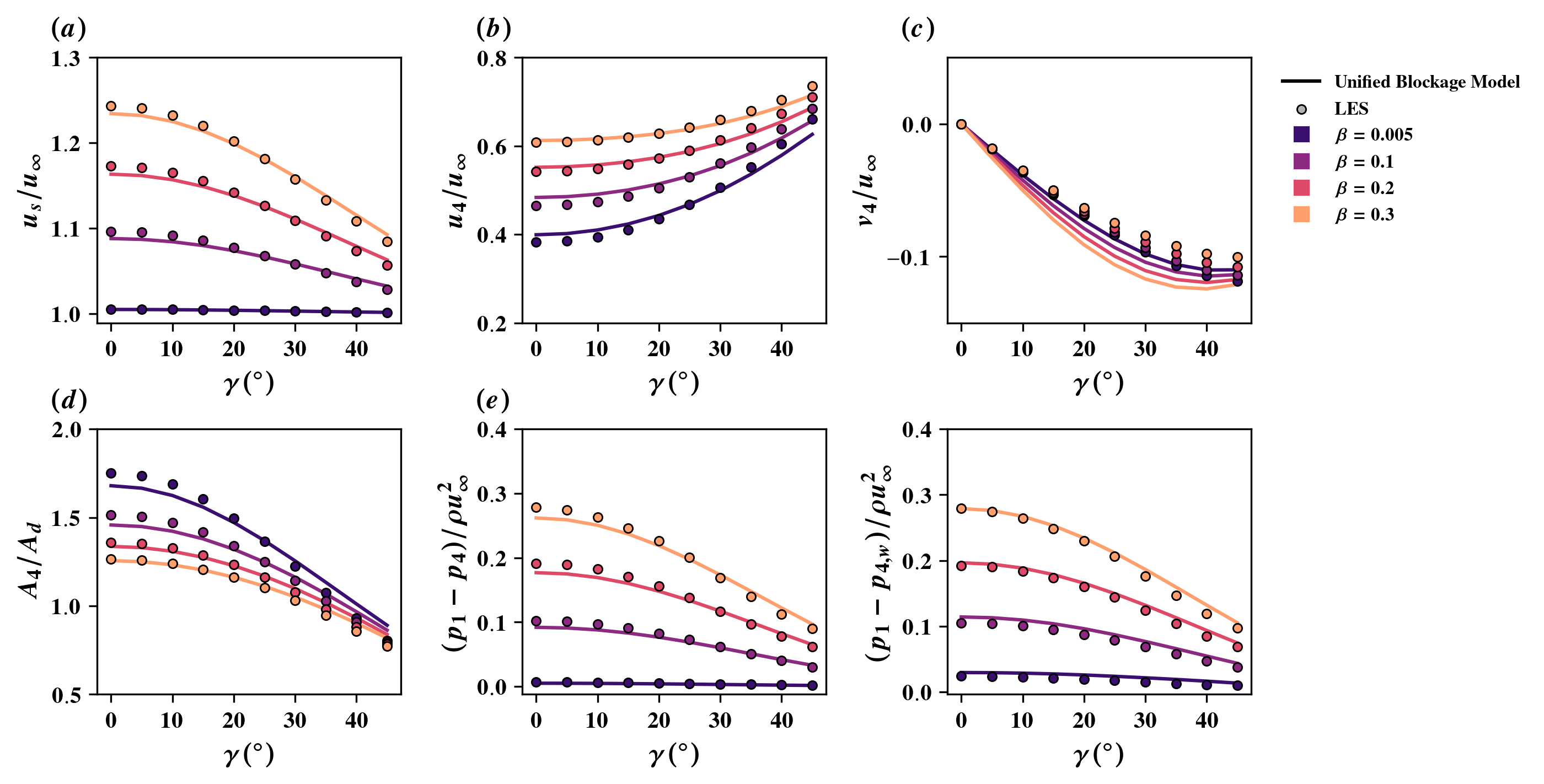}
	\caption{Comparison of (a) bypass velocity, (b) initial streamwise wake velocity, (c) spanwise initial wake velocity, (d) outlet wake area, (e) bypass pressure drop, and (f) streamtube pressure drop, between the Unified Blockage Model developed in~\S\ref{sec:model_derivation} and large eddy simulations for various misalignment angles.}
    \label{fig:yaw_streamtube_analysis}
\end{figure}

In figure~\ref{fig:yaw_streamtube_analysis}, the bypass velocity, streamwise and spanwise wake velocities, outlet wake area, and outlet pressures from LES and the Unified Blockage Model are shown for the misaligned actuator disk. 
Overall, the analytical model shows strong agreement with LES.
The actuator disk induces a lower bypass velocity as it becomes misaligned with the inflow because both the induction and projected area decrease (figure~\ref{fig:yaw_streamtube_analysis}($a$)).
The reduced induction also causes the streamwise wake velocity to increase with increasing blockage and magnitude of misalignment (figure~\ref{fig:yaw_streamtube_analysis}($b$)).
Figure~\ref{fig:yaw_streamtube_analysis}($c$) shows that the analytical model captures the overall magnitude of the spanwise wake velocity. However, the subtle variations with blockage are reversed relative to the LES predictions.
Misalignment reduces induction, causing the wake to expand less, as well as incurring a $\cos(\gamma)$ decrease in the projected area of the rotor, both of which reduce the wake area (figure~\ref{fig:yaw_streamtube_analysis}($d$)).
Figure~\ref{fig:yaw_streamtube_analysis}($e$) demonstrates that the blockage-induced pressure gradient becomes weaker with increasing magnitude of misalignment as a result of the reduced bypass velocity. This reveals that misalignment weakens the blockage effect, as will be discussed further. Finally, figure~\ref{fig:yaw_streamtube_analysis}($f$) shows that the wake pressure measured in LES is also well captured by the model.

\begin{figure}
    \centering
    \includegraphics[width=\linewidth]{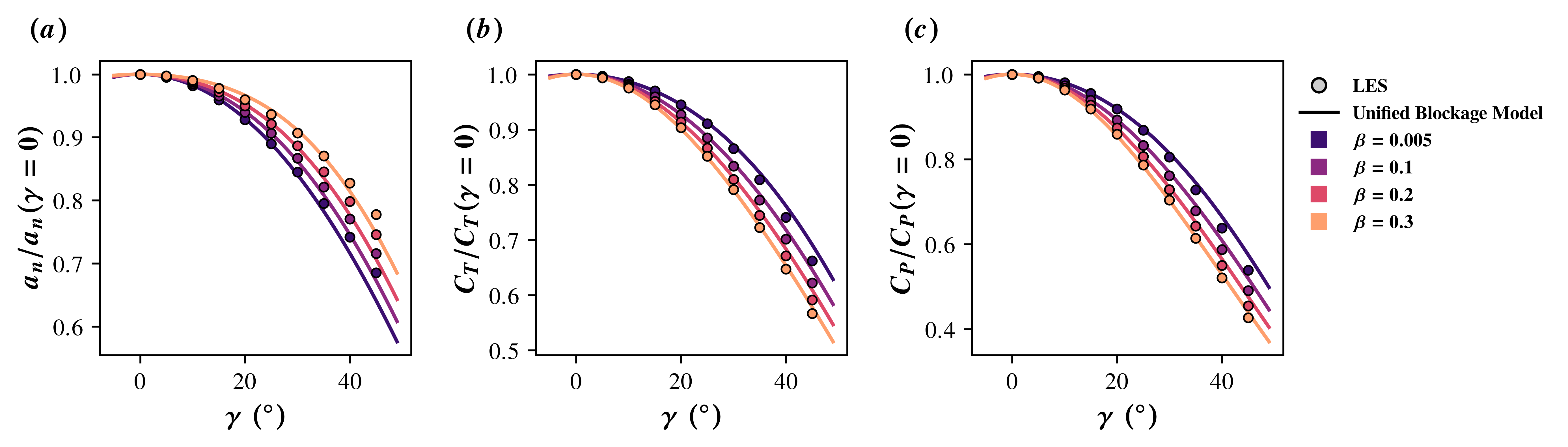}
    \caption{(a) Induction factor, (b) thrust coefficient, and (c) power coefficient normalised by their inflow-aligned ($\gamma=0$) values for varying blockage ratios. Unified Blockage Model predictions (Eqs.~\ref{eq:final_1}–\ref{eq:final_6}) and LES results are shown for $C_T' = 2$. Increasing blockage leads to a more gradual reduction in induction with misalignment, while thrust and power decrease more rapidly with misalignment.}
    \label{fig:tilt_sweep_normalized}
\end{figure}

In figure~\ref{fig:tilt_sweep_normalized}, we replot the induction, thrust and power of the misaligned rotor normalized by their inflow-aligned ($\gamma=0$) values for each blockage ratio $\beta$.
This reveals a central result of the present study: as the blockage ratio increases, the induction factor decreases more gradually with increasing misalignment angle. Correspondingly, as blockage increases, rotor thrust and power decrease more rapidly with increasing misalignment angle.
This behaviour arises because rotor misalignment reduces the magnitude of the blockage effect by decreasing both the projected rotor area and the rotor thrust. 
By lowering the thrust force, misalignment reduces induction. However, the reduced thrust weakens the blockage effect, which would otherwise act to reduce induction as well.
As a result, the decrease in induction with misalignment is more gradual when the rotor operates under confinement. The slower reduction in induction implies a lower disk velocity, which leads to a more rapid loss of thrust and power with misalignment when operating in a confined environment.
Therefore, the effects of blockage and misalignment are coupled through the turbine thrust force. Due to this coupling, existing models for misalignment and blockage cannot be superimposed. Instead, these effects must be modelled jointly as provided by the Unified Blockage Model.
In addition to implications of these results on measuring thrust and power for misaligned rotors in wind/water tunnels, these results may affect the future design of wake steering flow control for hydrokinetic turbines \cite[]{modali2021towards}.

\begin{figure}
    \centering
    \includegraphics[width=0.8\linewidth]{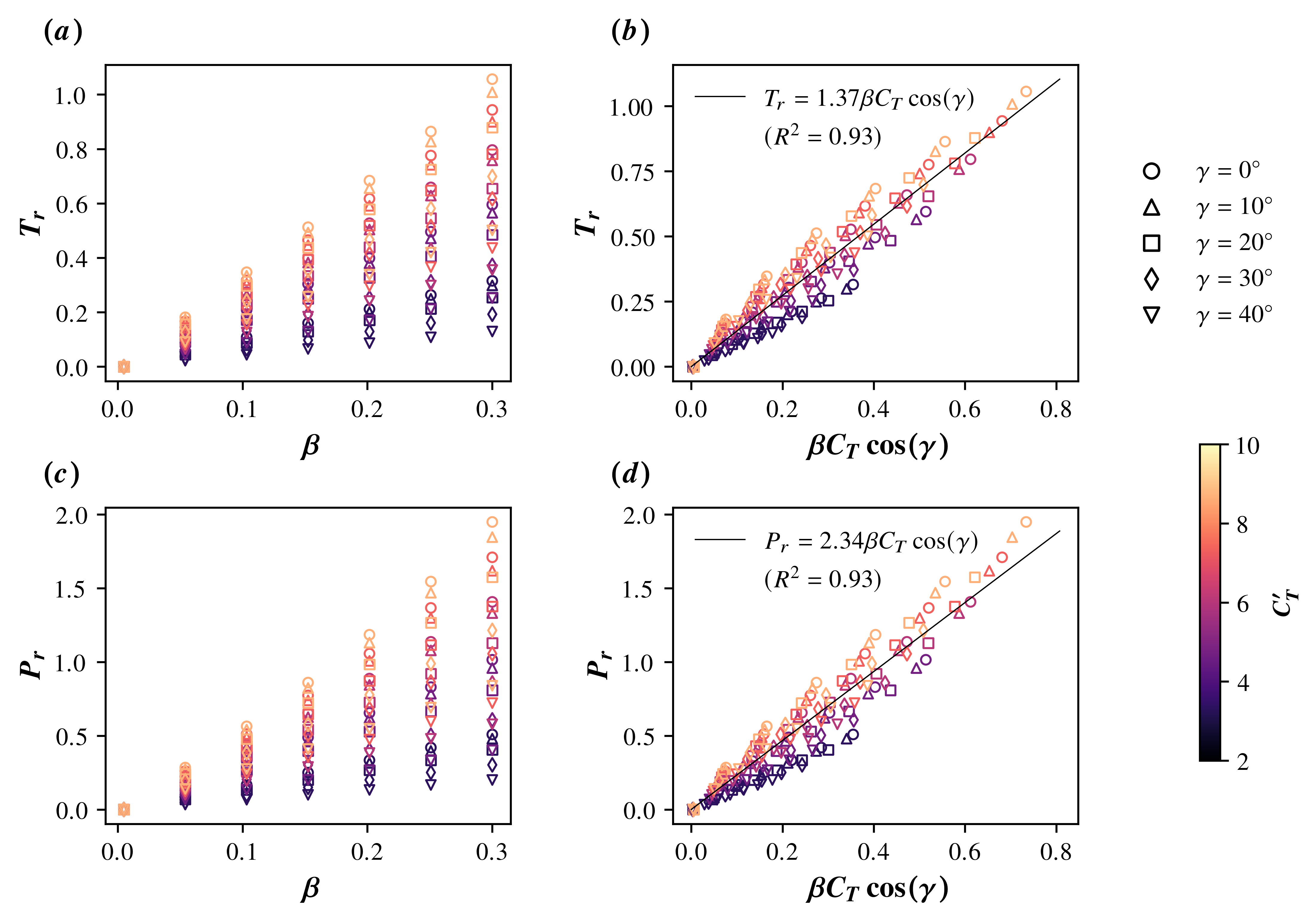}
    \caption{The blockage-induced change in thrust, quantified by $T_r$, is shown over $0 \le \gamma \le 40^\circ$ and $2 \le C_T' \le 10$ as a function of (a) $\beta$ and (b) $\beta C_T \cos(\gamma)$. Panels (c,d) show the corresponding results for power quantified by $P_r$.}
    \label{fig:blockage_effect_scaling}
\end{figure}

The influence of flow confinement on measured thrust (drag) and power is an important consideration in experimental fluid dynamics. 
A common approach is to characterise blockage effects using only the geometric blockage~\cite[]{schottler2018wind,upfal2025prf}. 
However, our results reveal that this is insufficient to characterise the magnitude of blockage effects, owing to its dependence on the thrust coefficient. 
Consequently, there is no universal blockage ratio below which blockage effects may be considered negligible, as this threshold depends on $C_T$.
We instead propose a more appropriate parameter based on the ratio of the streamwise component of the thrust force, $\vec{F}_T \cdot \hat{i}$, to the freestream momentum flux scale, $\tfrac{1}{2}\rho A_{CV} u_\infty^2$, yielding $\beta C_T \cos(\gamma)$. 
Note that in a confined wind/water tunnel setting, this $C_T$ is what would be measured in the experiment, so all information in $\beta C_T \cos(\gamma)$ would be known. 
To demonstrate this scaling, we quantify the magnitude of blockage effects as the change in thrust and power relative to an unblocked case with the same $C_T'$ and $\gamma$, through the thrust ratio $T_r = C_T/C_T(\beta=0) - 1$ and power ratio $P_r = C_P/C_P(\beta=0) - 1$.
Figure~\ref{fig:blockage_effect_scaling} demonstrates that, unlike $\beta$, the proposed ratio captures the leading-order scaling of the blockage-induced increases in thrust and power, quantified by $T_r$ and $P_r$, respectively. 
This simple scaling can be readily evaluated and provides a practical metric for assessing blockage effects in experimental configurations to yield a quick, first-order, quantitative estimate of the change in thrust and power due to blockage effects.

\subsection{Confined blade element momentum model}
\label{sec:results_bem} 
The Unified Blockage Model has been validated against LES of an actuator disk across a wide range of misalignment angles $\gamma$, local thrust coefficients $C_T'$, and blockage ratios $\beta$ in the previous section. 
However, the Unified Blockage Model is not fully predictive for a rotating turbine, since $C_T'$ depends on the aerofoil characteristics as well as the blade pitch angle $\theta_p$, tip-speed ratio $\lambda$, and yaw angle $\gamma$. Moreover, once $a_n$ is determined, the power of a rotating turbine still explicitly depends on the aerofoil properties and the control setpoints.
To address this, we develop a fully predictive Unified BEM model for a rotating turbine under arbitrary blockage and misalignment by coupling the Unified Blockage Model with a blade element model in \S\ref{sec:bem_model}. Here, we run the Unified BEM model using lift and drag polars of~\cite{wimshurst2016computational} for the RISO-A1-24 aerofoil with $Re = 12 \times 10^6$. A 3D stall delay correction was applied to the lift and drag coefficients of the aerofoil~\cite[]{du1998stall}. MITRotor~\cite[]{liew2025mitrotor}, an open source tool for physics-based rotor modelling and optimisation, is used to solve the BEM model described in \S\ref{sec:bem_model}. The model is solved using 40 radial grid points.

\begin{figure}
    \centering
    \includegraphics[width=0.75\linewidth]{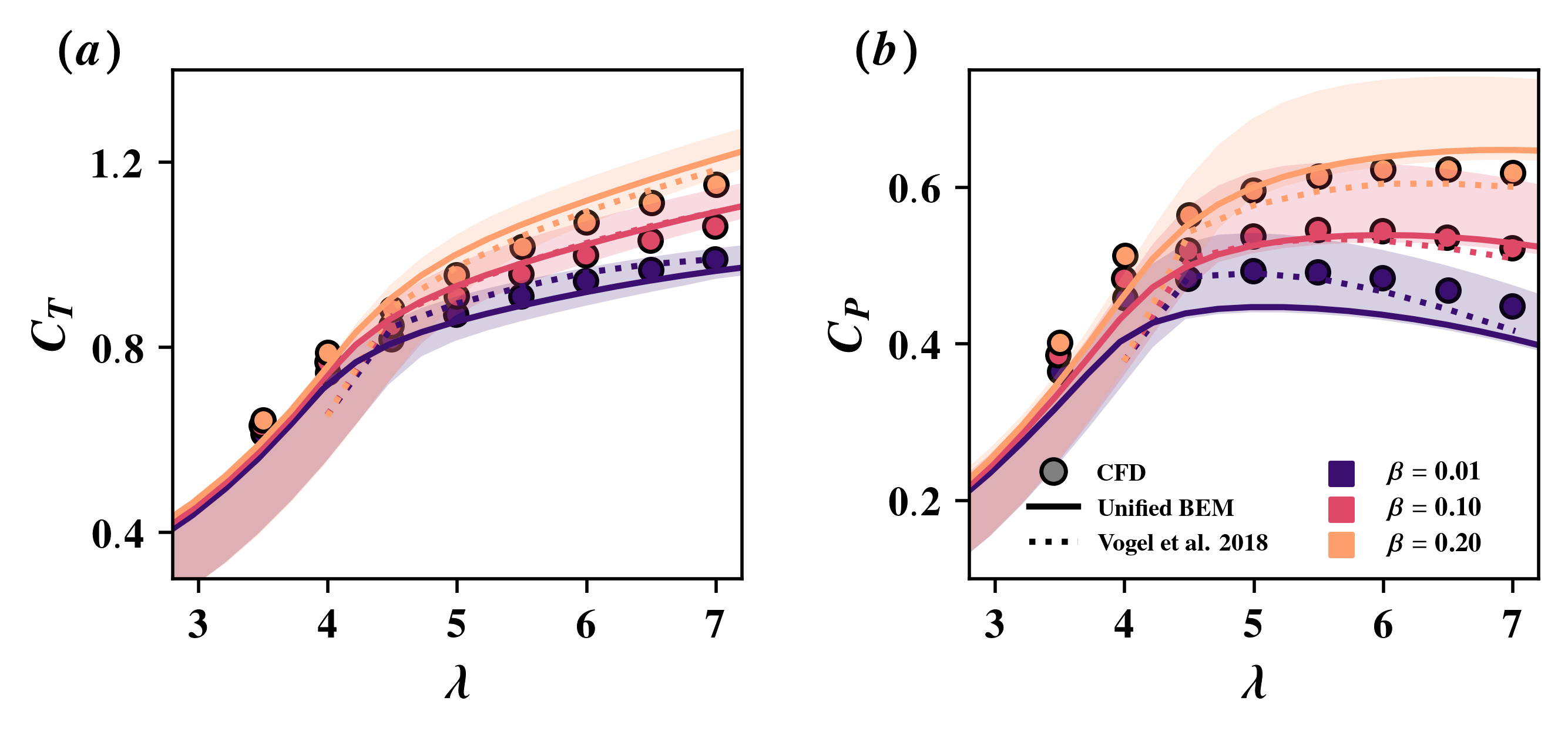}
    \caption{Validation of the Unified BEM method developed in~\S\ref{sec:bem_model} with blade-resolved simulations from~\cite{wimshurst2016computational}. The earlier BEM model results from \cite{vogel2018blade} are also shown for comparison. The shaded bands indicate the full spread of results obtained when the radial grid resolution is varied, and the tip loss, 3D stall delay, or tangential induction submodels are toggled.}
    \label{fig:BEM_Ct_Cp}
\end{figure}

In figure~\ref{fig:BEM_Ct_Cp}, the Unified BEM predictions are evaluated against the blade-resolved simulations of \cite{wimshurst2016computational}. 
The model reproduces the dependence of thrust and power on tip-speed ratio, and captures the systematic increase in both quantities with blockage. Importantly, the Unified BEM predictions capture the location of peak power across all blockage ratios considered.
In general, BEM models rely on blade element modelling to capture lift and drag forces, using two-dimensional aerofoil polars and empirical submodels for tip losses and three-dimensional stall-delay effects, both of which introduce uncertainty into the model predictions. 
These uncertainties are distinct from the momentum modelling described in the Unified Blockage Model (the momentum portion of BEM used here), and instead come from the blade element portion of BEM.
These uncertainties are illustrated by the shaded bands plotted in figure~\ref{fig:BEM_Ct_Cp}, encompassing the full spread of results obtained when the radial grid resolution is reduced to 20 points, and the tip loss, 3D stall delay, or tangential induction submodels are toggled on or off.
Quantifying the uncertainty associated with this reasonable range of modelling choices reveals substantial variability in the predictions and highlights the opportunity to improve existing engineering models for the aero/hydrodynamics of rotor blades.
Despite these limitations, the overall agreement between the BEM predictions and the blade-resolved simulations is good and comparable to that reported in previous BEM validation studies.
Figure~\ref{fig:BEM_Ct_Cp} further compares the Unified BEM results with the model of~\cite{vogel2018blade}, which uses the momentum model developed by \cite{garrett2007efficiency}. The two BEM models demonstrate similar levels of agreement with the blade-resolved simulations of~\cite{wimshurst2016computational}.
Notably, the Unified BEM method generalises to rotor-inflow misalignment and high thrust coefficient by leveraging the Unified Blockage Model. In the following section, we demonstrate that the errors and uncertainties inherent to blade-element modelling can be eliminated by mapping rotor measurements between blockage ratios using a blockage correction method. 


\subsection{Blockage correction method based on the Unified Blockage Model}
\label{sec:results_correction}

\begin{figure}
    \centering
    \includegraphics[width=0.75\linewidth]{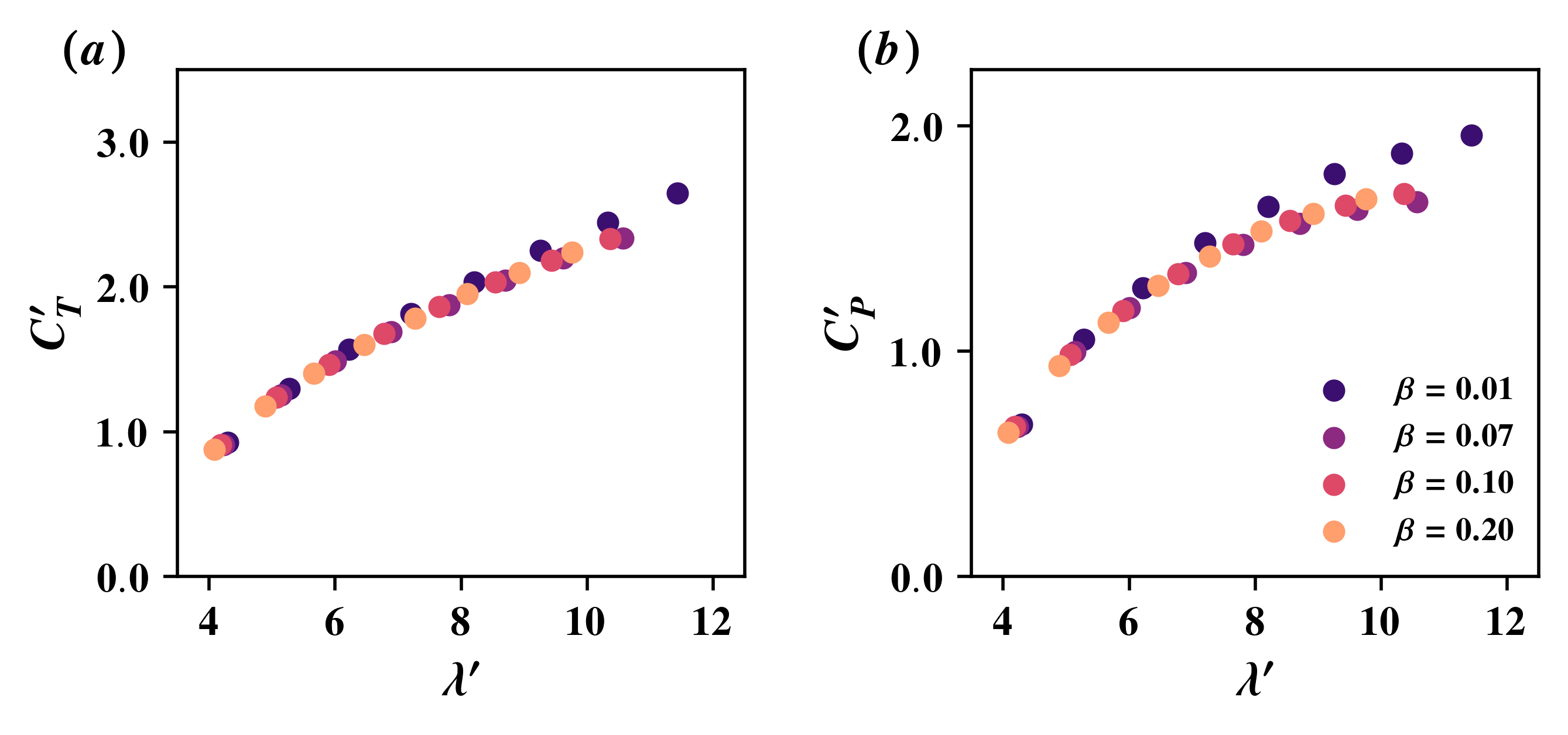}
    \caption{A scaling collapse of the (a) thrust and (b) power coefficients is demonstrated using the local parameters $\lambda'$, $C_T'$, and $C_P'$ for blade-resolved simulations from \cite{wimshurst2016computational} over a range of blockage ratios $\beta$. The observed collapse supports the validity of the blockage correction method presented in \S\ref{sec:correction_method}.}
    \label{fig:scaling_ct_cp}
\end{figure}

In section~\ref{sec:results_bem}, we validated a fully predictive rotor model for arbitrary misalignment angles, thrust coefficients, and blockage conditions against blade-resolved simulations. However, the reliance on two-dimensional aerofoil lift and drag polars, together with sub-models for tip losses and three-dimensional aerofoil corrections, introduces errors and  uncertainties into BEM-based predictions.
Alternatively, if measurements of rotor thrust and power are available at a single blockage ratio, these can be used instead of a blade element model to make predictions at arbitrary blockage ratios. In \S\ref{sec:correction_method}, we developed a new method that extracts the local thrust coefficient $C_T'$ and local power coefficient $C_P'$ as functions of the local tip-speed ratio $\lambda'$ using the Unified Blockage Model. 
This scaling, resulting in blockage-invariant quantities $C_T'(\lambda')$ and $C_P'(\lambda')$, describes the tip-speed, thrust and power of the rotor in relation to the rotor-normal disk velocity instead of the freestream velocity. 
In Appendix~\ref{sec:appendix_norm_bem}, we show that when the aerofoil lift and drag curves are independent of blockage, $C_T'$ and $C_P'$ then only depend on $\lambda'$, $\theta_p$, and $\gamma$. 
In this case, we expect such information to be sufficient for predicting the performance of a given rotor at other blockage ratios. 
In figure~\ref{fig:scaling_ct_cp}, we demonstrate the proposed scaling, $C_T'(\lambda')$ and $C_P'(\lambda')$, using the blade-resolved simulations of \cite{wimshurst2016computational}. 
This scaling requires knowledge of the turbine induction, which is computed using the Unified Blockage Model. 
In the lowest blockage case ($\beta = 0.01$), $C_T'$ and $C_P'$ are slightly higher than in the blocked cases. 
Despite this deviation in the unblocked case, the data collapses well overall. 
The collapse of the data reveals that these local parameters are constant with blockage ratio, supporting the argument presented in Appendix~\ref{sec:appendix_norm_bem}.




\begin{figure}
    \centering
    \includegraphics[width=0.9\linewidth]{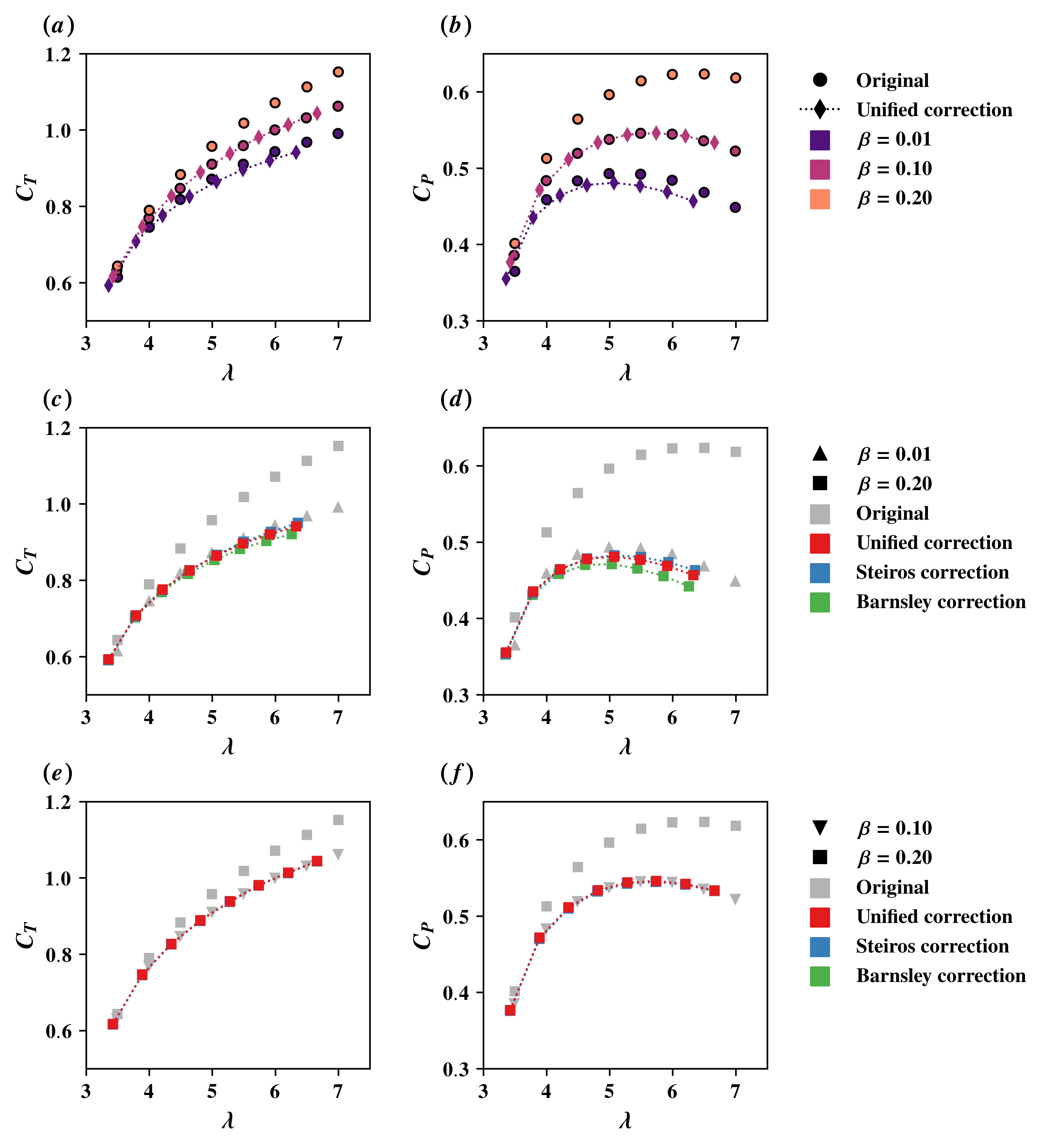}
    \caption{Blockage correction method developed in~\S\ref{sec:correction_method} using the Unified Blockage Model is applied to ($a$) thrust and ($b)$ power coefficients from blade-resolved simulations by~\cite{wimshurst2016computational}. The $\beta=0.2$ results are mapped to $\beta=0.1$ and $\beta=0.01$. ($c, d$) The present method is compared with the corrections of~\cite{steiros2022analytical} and~\cite{barnsley1990stall} to map the $\beta = 0.2$ case to $\beta=0.01$. ($e, f$) The present method is compared with the correction of~\cite{steiros2022analytical} to map the $\beta = 0.2$ case to $\beta=0.1$. The correction of~\cite{barnsley1990stall} is not shown as it only allows for mapping to $\beta = 0$.}
    \label{fig:blockage_correction}
\end{figure}

Having verified the proposed scaling, we now apply the blockage correction method developed in \S\ref{sec:correction_method} to the blade-resolved simulations of~\cite{wimshurst2016computational}. In figures~\ref{fig:blockage_correction}($a$) and \ref{fig:blockage_correction}($b$), the thrust and power coefficients, $C_T$ and $C_P$, from $\beta = 0.20$ are mapped to $\beta = 0.10$ and $\beta = 0.01$.
The blockage-corrected thrust and power coefficients show low error relative to the $\beta = 0.10$ case, reflecting the strong collapse of the blocked cases ($\beta = 0.07, 0.10, 0.20$) in figure~\ref{fig:scaling_ct_cp}. Larger errors occur when mapping to the $\beta = 0.01$ case, consistent with the slight divergence of $C_T'$ and $C_P'$ between blocked and unblocked conditions. This indicates that the remaining error may originate from blockage-induced changes in local rotor performance not accounted for by the correction.
In figures~\ref{fig:blockage_correction}($c$) and~\ref{fig:blockage_correction}($d$), we compare the present method with the corrections of~\cite{steiros2022analytical} and~\cite{barnsley1990stall} to map the $\beta = 0.20$ measurements to $\beta=0.01$. The present method performs similarly to~\cite{steiros2022analytical} while both methods yield notable improvements compared to \cite{barnsley1990stall} for higher $C_T$, achieved at larger values of the tip-speed ratio.
Figures~\ref{fig:blockage_correction}($e$) and~\ref{fig:blockage_correction}($f$) shows that the present method and the correction of~\cite{steiros2022analytical} both exhibit excellent agreement when mapping from $\beta = 0.20$ to $\beta = 0.10$. The method of \cite{barnsley1990stall} is not shown as it only allows for mapping to $\beta = 0$.

\begin{figure}
    \centering
    \includegraphics[width=0.9\linewidth]{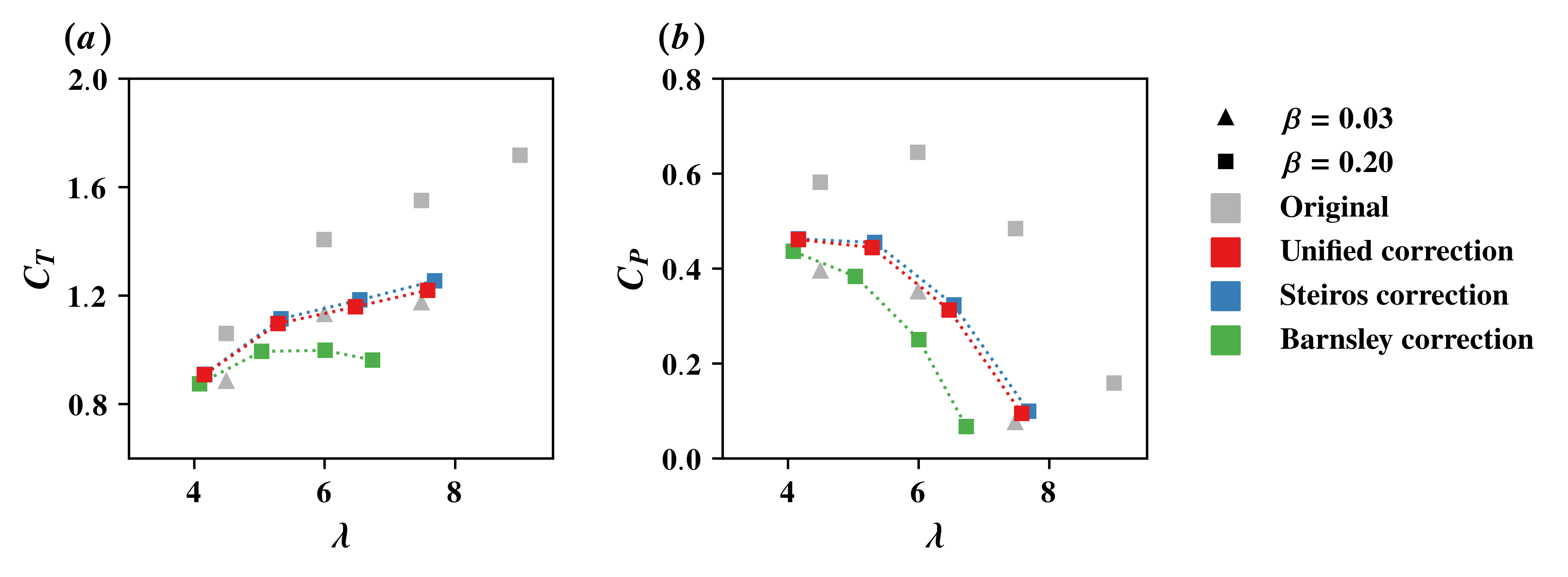}
    \caption{Blockage correction method developed in~\S\ref{sec:correction_method} using the Unified Blockage Model is applied to ($a$) thrust and ($b$) power coefficients from the actuator line simulations of~\cite{steiros2022analytical} and compared with the methods of~\cite{barnsley1990stall} and~\cite{steiros2022analytical}. The confined ($\beta=0.20$) case is mapped to the near-unconfined ($\beta=0.03$) data.}
    \label{fig:blockage_correction_steiros_data}
\end{figure}

In figure~\ref{fig:blockage_correction_steiros_data}, we apply the three corrections to the large eddy simulations of an actuator line model performed by~\cite{steiros2022analytical} for a high solidity (high $C_T$) rotor. The corrections are used to map the confined ($\beta = 0.20$) thrust and power coefficients to the unconfined ($\beta = 0.03$) results. The three corrections perform similarly at lower tip speed ratios, but the method of~\cite{barnsley1990stall} shows significant error at higher tip-speed ratios, corresponding to high thrust coefficients where classical momentum theory is no longer valid. The present model, which accurately captures the dynamics of high thrust rotors, exhibits lower error across all tip speed ratios.

\begin{figure}
    \centering
    \includegraphics[width=0.9\linewidth]{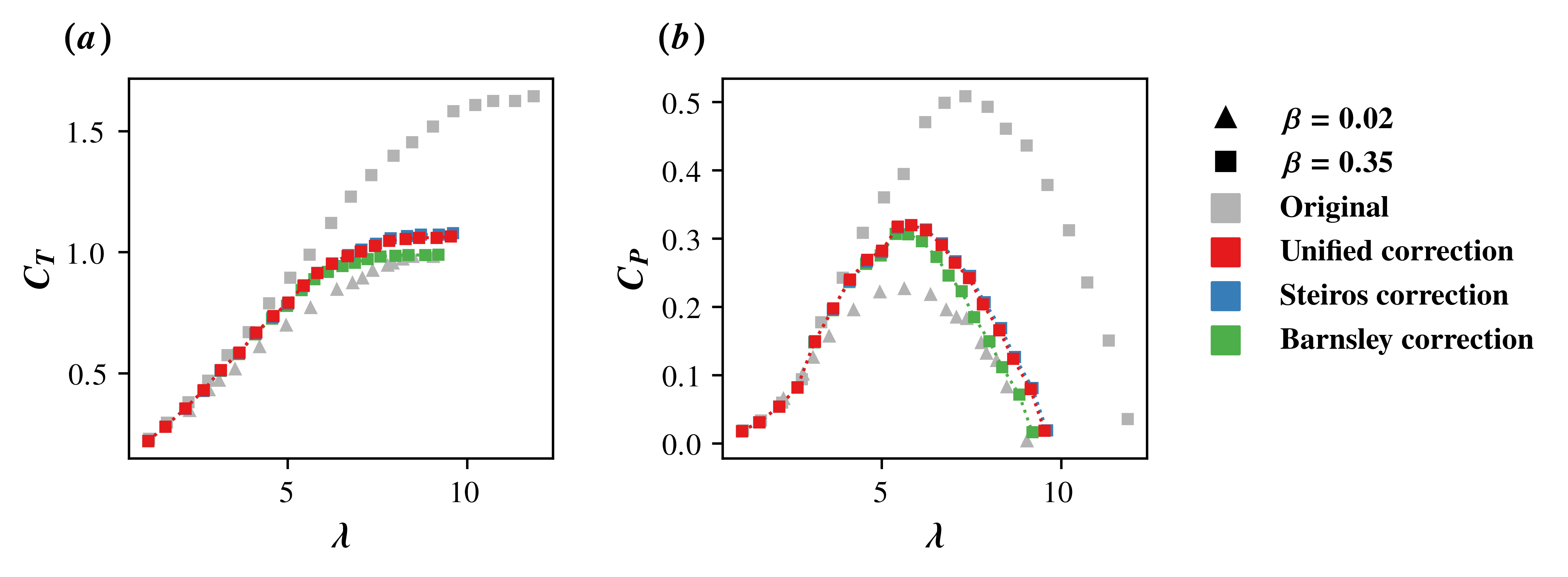}
    \caption{Blockage correction methods applied to the (a) thrust and (b) power coefficients from the ~\cite{ross2020experimental} dataset. The confined case ($\beta = 0.35$) case is mapped to the near-unconfined case ($\beta = 0.02$).}
    \label{fig:ross_correction}
\end{figure}

Finally, we evaluate the blockage correction methods against experimental data. 
The experimental dataset of~\cite{ross2020experimental} compares the power and thrust of identical axial flow hydrokinetic turbines at two blockage ratios, $\beta=0.02$ and $\beta = 0.35$. 
Figure~\ref{fig:ross_correction} shows the blockage correction developed in~\S\ref{sec:correction_method} as well as the~\cite{steiros2022analytical} and~\cite{barnsley1990stall} methods applied to these measurements. All the corrections show similar discrepancies compared with the low blockage reference data, particularly for $C_P$. 
The lower agreement for $C_P$ is also reported in \cite{ross2020experimental}.
The relatively large discrepancy in corrected $C_P$ differs notably from the more accurate corrected $C_P(\lambda)$ curves using simulation data shown previously (figures~\ref{fig:blockage_correction} and~\ref{fig:blockage_correction_steiros_data}). 

\begin{figure}
    \centering
    \includegraphics[width=0.8\linewidth]{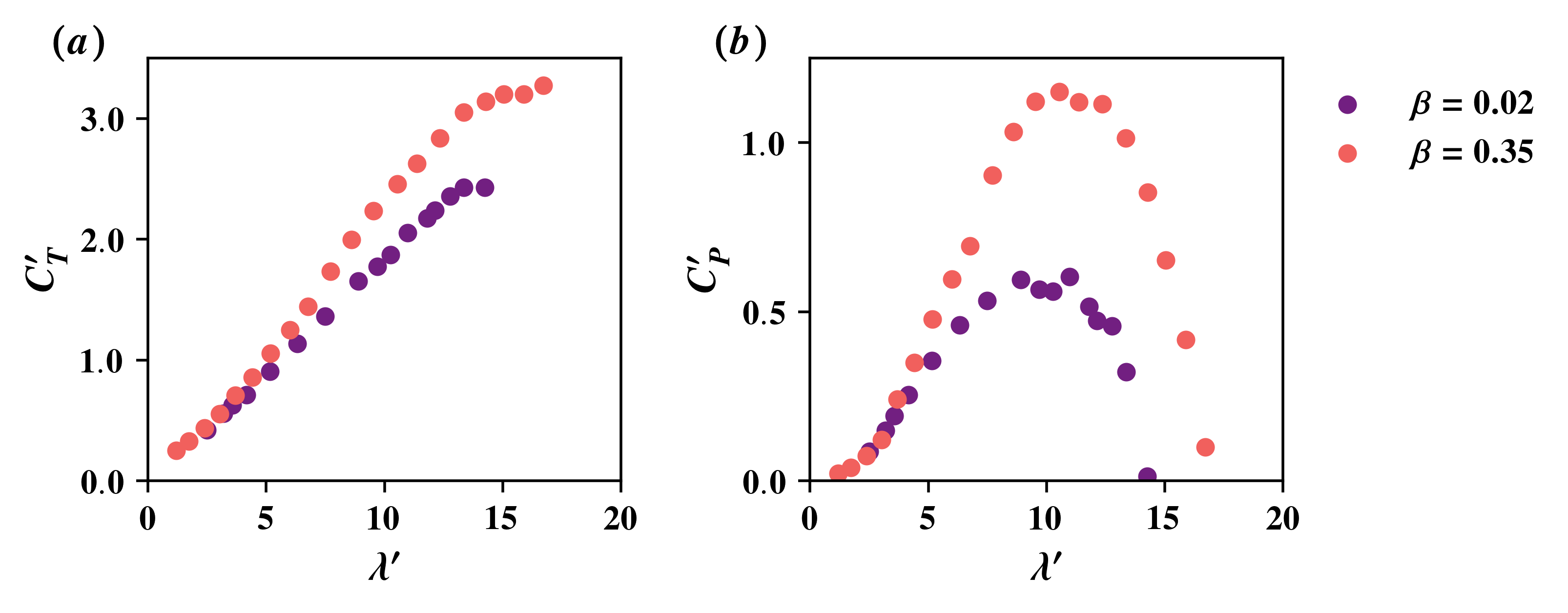}
    \caption{Scaling of the~\cite{ross2020experimental} measurements taken at two blockage ratios using the disk velocity computed with the Unified Blockage Model. The results show that ($a$) the local thrust coefficient and ($b$) the local power coefficient do not collapse with $\lambda'$.}
    \label{fig:ross_scaling_beta}
\end{figure}

\begin{figure}
    \centering
    \includegraphics[width=0.8\linewidth]{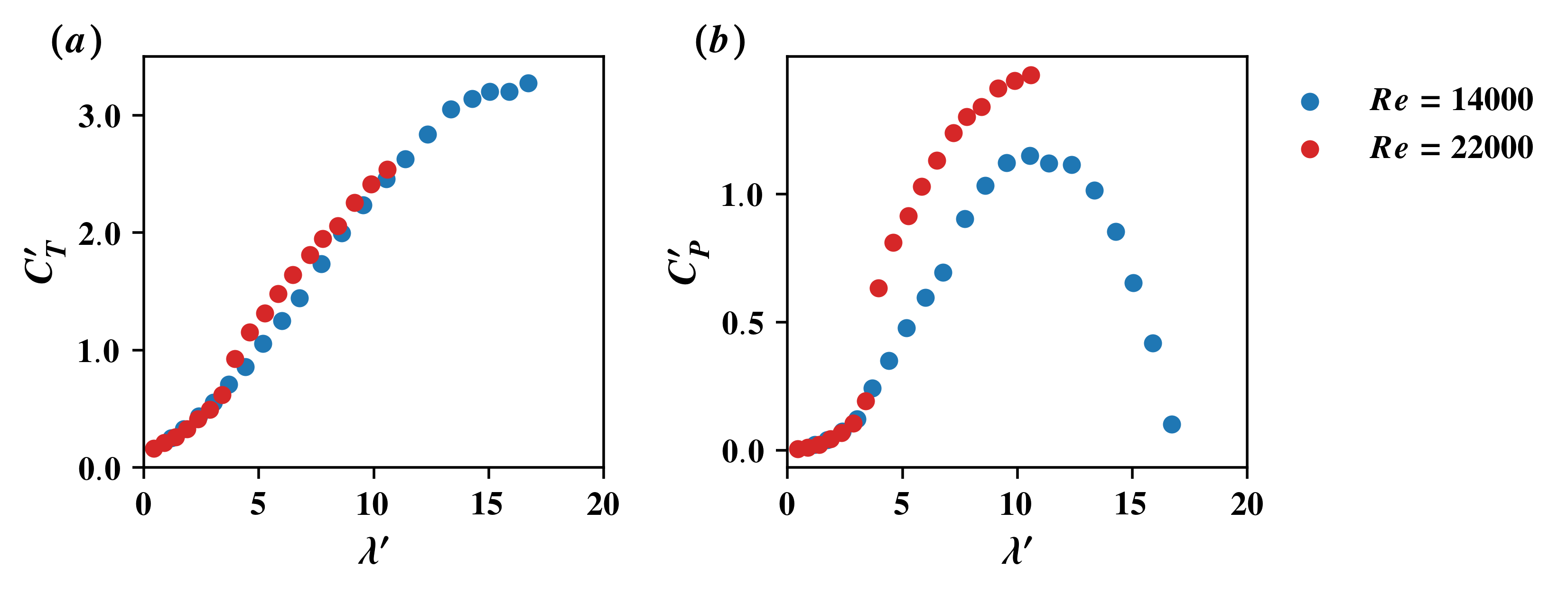}
    \caption{Scaling of the ($a$) thrust and ($b$) power measurements from~\cite{ross2020experimental} taken at two Reynolds numbers using the disk velocity computed with the Unified Blockage Model, revealing changes in local rotor performance with Reynolds number.}
    \label{fig:ross_scaling_Re}
\end{figure}

To elucidate challenges with blockage corrections in the experimental dataset, we show the scaling to $C_T'(\lambda')$ and $C_P'(\lambda')$ based on the local rotor velocity that underpins the blockage correction model from \S\ref{sec:correction_method}.
Figure~\ref{fig:ross_scaling_beta} demonstrates that, unlike the results of~\cite{wimshurst2016computational}, the \cite{ross2020experimental} dataset does not collapse under the $C_T'(\lambda')$ and $C_P'(\lambda')$ scaling,
suggesting that the local rotor performance may be changing with $\beta$. 
This is acknowledged by~\cite{ross2020experimental}, who note that blade-level Reynolds number independence was not maintained across their tests.
Specifically, \citet{ross2020experimental} noted that rotor performance was sensitive to changes in chord-based Reynolds number. 
Using their $C_T(\lambda)$ and $C_P(\lambda)$ curves at two chord-based Reynolds numbers, we perform the scaling to $C_T'(\lambda')$ and $C_P'(\lambda')$.
As shown in figure~\ref{fig:ross_scaling_Re}, these measurements also fail to collapse under the proposed scaling, highlighting that the measurements were conducted under conditions where the local rotor performance---and consequently the lift and drag characteristics of the blades---changes with Reynolds number. Since the induction changes with blockage, blockage affects the chord-based Reynolds number, leading to different rotor performances when the blockage is changed.

This Reynolds-number sensitivity violates a key assumption of the blockage correction method, namely that the aerofoil lift and drag curves remain invariant with $\beta$. 
More generally, we expect that any turbine blockage correction approach 
will fail when trying to correct data that spans a significant aero/hydrofoil Reynolds number dependency unless those dependencies are explicitly and accurately captured in the blockage correction model.
This analysis further reveals the need for experimental measurements of identical turbines at multiple blockage ratios and sufficiently high Reynolds number to achieve Reynolds-number independence. Such a dataset would be highly valuable to validate engineering models for rotors operating under blockage.

\FloatBarrier

\section{Conclusions}
\label{sec:conclusions}
In confined flows, mass conservation requires acceleration of fluid around the rotor, which induces a favourable pressure gradient.
This pressure gradient alters the balance of streamwise momentum, increasing the rotor disk velocity relative to unblocked conditions. 
Consequently, blockage reduces rotor induction while increasing thrust and power. Because the turbine thrust itself generates the blockage-induced pressure gradient, there is a coupled feedback between the thrust force and blockage effects. 

Rotor misalignment reduces the rotor-normal disk velocity, lowering the turbine thrust and consequently reducing induction. 
When a rotor is misaligned in confined flow, misalignment and blockage effects are coupled through the thrust force: misalignment reduces the effective blockage by lowering both the thrust force and the projected rotor area.
This mechanism causes the induction factor to decrease more gradually with misalignment in confined flows.
As a result, the thrust and power reductions associated with misalignment are exacerbated in confined conditions. 
Since the misalignment and blockage effects are coupled, they must be modelled jointly as the present approach provides.

This work develops a Unified Blockage Model to predict the induced velocities of an actuator disk across arbitrary blockage ratios, misalignment angles, and thrust coefficients. The analytical model is validated against LES, showing excellent agreement with the high-fidelity simulations and lower error than existing models. The model also accurately captures the high-thrust, low-blockage limit by including the base suction pressure deficit from the Unified Momentum Model for unconfined rotors \cite[]{liew2024unified}.

Since the thrust coefficient for an arbitrary rotor design is not known a priori, the Unified Blockage Model is not fully predictive for a rotating turbine.
Therefore, a Unified BEM model is developed by coupling the validated momentum model with a blade-element closure for the thrust coefficient.
This yields a fully-predictive rotor model for arbitrary blade design, blockage ratio, and control settings.
The resulting framework is validated against blade-resolved computational fluid dynamics simulations.

When measurements of rotor thrust and power are available, these can be used to extract the local thrust and power coefficients as functions of the local tip-speed ratio. We identify that if the blade lift and drag properties are independent of blockage, then the local rotor performance is also independent of blockage. This novel scaling is verified with blade-resolved simulation results.
We then use this identified scaling to develop a new blockage correction method for mapping measurements between arbitrary blockage ratios. The method is validated against blade-resolved simulations, actuator line simulations, and experimental measurements of an axial flow turbine. 
While the blockage correction method yields low errors when correcting simulation data, the blockage correction performs less well for existing experimental data due to the lack of Reynolds number independence across the different experiments at different blockage ratios.
Future work may consider experimental measurements that vary the blockage ratio while ensuring Reynolds number independence.
The novel blockage correction method can be used to predict rotor performance in untested blockage ratios without introducing the uncertainties associated with blade element models.

The Unified Blockage Model captures the coupled effects of blockage, misalignment, and arbitrary thrust coefficient on the induced velocities of an actuator disk. 
Based on this model, the new blockage correction provides a practical tool to account for blockage effects in turbine and propeller measurements. 
Finally, the Unified BEM method can be used to optimise the design and control of rotors operating in confined flows, such as hydrokinetic turbines in shallow water or wind turbines in shallow atmospheric boundary layers with stable capping inversions.

\section*{Acknowledgements}
The authors acknowledge funding from the National Science Foundation (Fluid Dynamics program, grant number FD-2542240).
In addition, K.S.H. acknowledges funding through a National Science Foundation Graduate Research Fellowship under grant no. DGE-2141064. 
Simulations were performed on Stampede3 supercomputers under the NSF ACCESS project ATM170028.

\section*{Declaration of Interests}
The authors report no conflict of interest.
\appendix

\section{Base suction pressure from Unified Momentum Model}
\label{sec:appendix_umm}
The Unified Momentum Model (UMM) describes the induction of an actuator disk operating in unconfined flow ($\beta\rightarrow0$) for arbitrary thrust coefficient and misalignment angle~\cite[]{liew2024unified}. 
The UMM captures the high thrust regime of rotors by modelling the base suction effect, which is a persistent pressure drop in the near-wake region. 
In confined flow, a uniform pressure gradient develops which spans the full channel cross-section and dominates over the contribution of the base suction pressure deficit.
In the present work, we include the base suction pressure deficit predicted by the UMM to ensure high accuracy across arbitrary blockage levels, misalignment angles and thrust coefficients.

The UMM equations solve for the rotor-normal induction $a_n$, streamwise wake velocity $u_4$, lateral wake velocity $v_4$, near-wake length $x_0$, and the pressure difference $(p_{4,w}-p_1)$. Note that in the context of UMM, $p_{4,w}$ refers to the pressure within the streamtube at $x=x_0$.
The final form of the UMM equations is
\begin{align}
\label{eq:final_1_umm}
a_n &= 1 - \sqrt{\frac{u_\infty^2 - u_4^2 - v_4^2}{C_T^\prime \cos^2(\gamma) u_\infty^2}-\frac{(p_{4,w} - p_1)}{\frac{1}{2} \rho C_T^\prime \cos^2(\gamma) u_\infty^2}}\\ \label{eq:final_2_umm}
u_4 &= -\frac{1}{4} C_T^\prime (1-a_n) \cos^2(\gamma)u_\infty + \frac{u_\infty}{2} + \notag \\ 
&\frac{1}{2}\sqrt{\left(\frac{1}{2} C_T^\prime (1-a_n) \cos^2(\gamma)u_\infty - u_\infty \right)^2 - \frac{4(p_{4,w} - p_1)}{\rho}}\\ 
v_4 &= -\frac{1}{4} C_T^\prime (1-a_n)^2 \sin(\gamma) \cos^2(\gamma) u_\infty\\ \label{eq:final_4_umm}
\frac{x_0}{D} &=\frac{\cos(\gamma)}{2\beta_s}\cfrac{u_\infty + u_4}{|u_\infty - u_4|}\sqrt{\frac{(1-a_n)\cos(\gamma)u_\infty}{u_\infty + u_4}} \\
\label{eq:final_5_umm}
p_{4,w} - p_1 &= -\frac{1}{2\pi} \rho C_T^\prime (1-a_n)^2 \cos^2(\gamma) u_\infty^2 \arctan\left[\frac{1}{2}\frac{D}{x_0}\right] + p^{NL}(C_T', \gamma, a_n, x_0),
\end{align}
where the unknown shear layer growth rate parameter is $\beta_s=0.1403$, based on the analysis of fundamental turbulent jet flow.
The nonlinear pressure field $P^{NL}(x,y)$ as a function of streamwise $x$ and spanwise $y$ directions is computed as
\begin{equation}
    P^{NL}(x, y) = \frac{\rho}{2\pi}\left(g_x \circledast \frac{x}{x^2 + y^2} +  g_y \circledast \frac{y}{x^2 + y^2}\right),
    \label{eq:pressure_convolution}
\end{equation}
where $\circledast$ is the two-dimensional convolution operator with advection terms
\begin{align}
    \label{eq:nonlinear_gx}
    g_x &= -\left(w_x\frac{\partial w_x}{\partial x} + w_y\frac{\partial w_x}{\partial y}\right)\\
    \label{eq:nonlinear_gy}
    g_y &= -\left(w_x\frac{\partial w_y}{\partial x} + w_y\frac{\partial w_y}{\partial y}\right).
\end{align}
The velocity field is decomposed into background flow and actuator disk-induced component $u=u_\infty + w_x$ and $v=w_y$.
There is no freestream flow $v_\infty$ in the lateral direction.
The velocity components are updated using the inviscid Euler equations for the nonlinear induced velocities. This system is solved by iteration and evaluated at $P^{NL}(x=x_0, y=0)$ which is written as $p^{NL}(C_T', \gamma, a_n, x_0)$.
The converged solution to the base suction pressure in Eq.~\eqref{eq:final_5_umm} gives the $p_\textrm{UMM}(C_T', \gamma)$ term used to close the Unified Blockage Model.

\section{Local normalisation of blade element equations}
\label{sec:appendix_norm_bem}
Here we show that under a specific set of assumptions, we expect the local thrust and power coefficients $C_T'$ and $C_P'$ to depend only on $\lambda'$, $\theta_p$ and $\gamma$, and are therefore independent of the blockage ratio $\beta$. For this analysis, we neglect tangential induction, assume $a_n$ to be constant over the rotor, and assume the aerofoil lift $C_l(\mu, \alpha)$ and drag $C_d(\mu, \alpha)$ are independent of $\beta$. These assumptions constrain the inflow velocity at the rotor and the aerofoil performance.

We begin by normalising the axial and tangential relative velocities (Eqs.~\ref{eq:BEM-start} and~\ref{eq:bem_vt}) by the rotor-normal disk velocity $\vec{u}_d \cdot \hat{n} = (1 - a_n) \cos(\gamma) u_\infty$,
\begin{align}
    \frac{v_n'(\mu,\psi)}{u_\infty} & = \frac{v_n(\mu,\psi)}{(1-a_n)\cos(\gamma)u_\infty} = 1                  \\
    \frac{v_t'(\mu,\psi)}{u_\infty} &= \frac{v_t(\mu,\psi)}{(1-a_n)\cos(\gamma)u_\infty}= \lambda' \mu - \cos(\psi)\tan(\gamma)
\end{align}
which now depend only on $\lambda'$ and $\gamma$, removing dependence on $a_n$ and therefore also removing dependence on $\beta$. Using Eqs.~\ref{eq:bem_w} and~\ref{eq:bem_phi} the relative flow speed and angle become,
\begin{align}
    \frac{w'(\mu, \psi)^2}{u_\infty^2} &= \frac{w(\mu, \psi)^2}{(1 - a_n)^2 \cos^2(\gamma) u_\infty^2} = 1 + (\lambda' \mu - \cos(\psi)\tan(\gamma))^2 \\
    \phi &= \tan^{-1} \left(\frac{1}{\lambda' \mu - \cos(\psi)\tan(\gamma)}\right).
\end{align}

Finally, we can show that $C_{n}$ and $C_{tan}$ no longer depend on $a_n$ or $\beta$. Since $\alpha(\mu,\psi) = \phi(\mu,\psi) - \theta_t(\mu) - \theta_p$ and $\phi$ is a function of $\lambda'$ and $\gamma$ only, then $C_{n}(\mu, \alpha)$ and $C_{tan}(\mu, \alpha)$ depend only on $\lambda'$, $\theta_p$, and $\gamma$.

Using the above expressions, we write the rotor-averaged local thrust coefficient as
\begin{equation}
    \overline{C}'_T = \frac{1}{\pi} \int_0^1 \int_0^{2\pi} \mu \sigma C_{n} w'(\mu, \psi)^2 d\mu d\psi
\end{equation}
and rotor-averaged local power coefficient as 
\begin{equation}
    \overline{C}'_P = \frac{1}{\pi} \int_0^1 \int_0^{2\pi} \mu^2 \lambda' \sigma C_{tan} w'(\mu, \psi)^2 d\mu d\psi
\end{equation}
which no longer depend explicitly on $a_n$ and therefore are also independent of $\beta$.

This analysis demonstrates that $C_T'$ and $C_P'$ depend only on $\lambda'$, $\theta_p$, and $\gamma$, and are independent of both $a_n$ and the blockage ratio $\beta$. Consequently, $C_T'$ and $C_P'$ are identical between flows with different blockage ratios when $\lambda'$, $\theta_p$, and $\gamma$ are matched, enabling the scaling proposed in \S\ref{sec:correction_method}.


\FloatBarrier
\bibliographystyle{jfm}
\bibliography{references}
\end{document}